\documentstyle[12pt,epsfig]{article}
\def \sech{\mathop{\rm sech}\nolimits}
\def \arctanh{\mathop{\rm arctanh}\nolimits}

\newcommand{\beq}{\begin{eqnarray}}
\newcommand{\eeq}{\end{eqnarray}}
\newcommand{\eqn}{\begin{equation}}
\newcommand{\een}{\end{equation}}
\begin{document}
\title{Kinks from Dynamical Systems: Domain Walls in a Deformed $O(N)$ Linear
Sigma Model}
\author{\dag A. Alonso Izquierdo, \dag M.A.
Gonz\'alez Le\'on \\ and \ddag J. Mateos Guilarte \\ \dag
Departamento de Matem\'atica Aplicada\\ \ddag Departamento de
F\'{\i}sica \\ Universidad de Salamanca, SPAIN}
\date{}
\maketitle


\begin{abstract}
It is shown how a integrable mechanical system provides all the
localized static solutions of a deformation of the linear
$O(N)$-sigma model in two space-time dimensions. The proof is
based on the Hamilton-Jacobi separability of the mechanical
analogue system that follows when time-independent field
configurations are being considered. In particular, we describe
the properties of the different kinds of kinks in such a way that
a hierarchical structure of solitary wave manifolds emerges for
distinct $N$.
\end{abstract}

\section{Introduction}

We divide this Introduction into three parts: A. A brief history
and the \lq\lq state of the  art". B. New developments and results
to be presented in this work . C. Scenarios of possible physical
applications.

\medskip

\noindent{\bf A}.

Kinks are solitary (non-dispersive) waves arising in several
one-dimensional physical systems. Here, we shall focus on the
relativistic theory of $N$-interacting scalar fields built on a
space-time that is the (1+1)-dimensional Minkowski space ${\bf
R}^{1,1}$. In this context, kinks are finite energy solutions of
the Euler-Lagrange equations, such that the time-dependence is
dictated by the Lorentz invariance:
$\vec{\phi}_K(x,t)=\vec{q}_K\left(
 \frac{x-vt}{\sqrt{1-v^2}}\right)$. Thus, the
search for kinks leads to the solving of a system of $N$-coupled
non-linear ordinary differential equations and therefore becomes a
very interesting problem in Mathematical Physics.

The study of topological defects began as an area of research in
field theory by the mid-seventies; see \cite{Co}. It was
immediately recognized that defects of the kink type are in
one-to-one correspondence with the separatrix trajectories between
the bounded and unbounded motion of a mechanical system, for which
the motion equations precisely form the non-linear system of
differential equations mentioned above. The equivalent mechanical
system is also Lagrangian and thus automatically integrable if
$N=1$. For $N\geq 2$, complete integrability is generically
non-guaranteed and the equivalence to a mechanical system is not
useful. This circumstance has been emphasized by Rajaraman; see
\cite{Ra} pp. 23-24, and partially circumvented by himself: the
trial orbit method allows one to guesstimate particular types of
kink trajectories.

There are, nevertheless, theories with $N=2$-coupled scalar fields
such that the equivalent dynamical system is completely
integrable. The prototype of this kind of system is the MSTB
model: in \cite{3} this was proposed in the context of the search
for non-topological solitons with stability provided by a $U(1)$
internal symmetry. In Reference \cite{4}, the model was considered
as a classical continuum approximation to a 1D crystal with a
two-component order parameter and it was shown that the search for
kinks in this system requires that a completely integrable
dynamical system be addressed. Ito, in a seminal paper \cite{5},
showed the Hamilton-Jacobi separability of the system of
non-linear differential equations. He found all the kink
trajectories and explained a very peculiar kink energy sum rule.

Very rich manifolds of kinks were discovered in two $N=2$ field
theoretical models, close relatives to the MSTB system, in a
recent research performed by the authors of the present work
\cite{6}. The investigation of kink properties in these models
requires the analysis of the separatrix trajectories in two
related dynamical systems which are type I and III respectively
in the classification of Liouville bidimensional $(N=2)$
completely integrable systems, see \cite{7}.

In fact, on choosing between the four types of Liouville dynamical
systems those that meet appropriate critical point structure, one
builds an enormous list of related $N=2$ field theoretical models
exhibiting manifolds of kinks of growing complexity; see \cite{8}.
The r\^ole of these models can be understood by noticing that the
MSTB system is a deformation of the $O(2)$-linear sigma model.
Instead of spontaneous symmetry breaking of $O(2)$ by a
degenerated $S^1$ vacuum manifold, the $O(2)$ symmetry group is
explicitly broken to ${\bf Z}_2\times {\bf Z}_2$ by a mass term;
only invariance under $\phi_a \to (-1)^{\delta_{ab}}\phi_a$, for
$b=1,2$, survives. From the point of view of quantum field theory,
this deformation is very natural because in (1+1)-dimensions
infrared divergences forbid the existence of Goldstone bosons,
according to a theorem of Coleman \cite{9}. Even if it is absent
in the classical action, a mass term will be generated by quantum
corrections.

We interpret this as follows: in the parameter space of the $N=2$
relativistic scalar field theories invariant under the ${\bf
Z}_2\times {\bf Z}_2$ with generators mentioned above, and
potential energy of the form
\[
U(\vec{\phi})\, =\, \frac{1}{2} \left( \alpha_1 \,
\phi_1^2+\alpha_2 \, \phi_2^2+\frac{\beta_1}{2}\,
\phi_1^4+\beta_{12}\,  \phi_1^2 \phi_2^2+ \frac{\beta_2}{2}\,
\phi_2^4\right) + C
\]
there are at least two distinguished points. There is a choice of
coupling constants such that there is explicit $O(2)$ symmetry,
which is spontaneously broken. This is the linear $O(2)$-sigma
model. The other interesting point is the MSTB model where the
explicit ${\bf Z}_2\times {\bf Z}_2$ symmetry generated by $\phi_a
\to (-1)^{\delta_{ab}}\phi_a$, for $b=1,2$, breaks spontaneously
to the ${\bf Z}_2$ sub-group generated by $\phi_2\to -\phi_2$. The
key observation is that the renormalization group flow induced by
quantum corrections in the parameter space avoids the $O(2)$-sigma
system and instead leads to the MSTB model, which also offers a
variety of kinks. All the other field theoretical models
exhibiting an abundant supply of kinks also correspond to
deformations of $O(2)$-symmetric systems with potential energies
that depend on higher powers of $\phi_1$ and $\phi_2$, \cite{6},
\cite{8}.

There are strong analogies with the Zamolodchikov $c$-theorem,
\cite{10}: deformations in the space of (1+1)-dimensional field
theories leading from conformal to integrable systems are the most
interesting ones. We meet an analogous finite dimensional
situation: replace the (infinite dimensional) conformal group by
the $O(2)$ group and integrability of one system with infinite
degrees of freedom by integrability of a bidimensional mechanical
system.

\medskip

\noindent {\bf B}.

This paper is devoted to investigating the kink solitary waves of
the deformation of the linear $O(N)$-sigma model that generalize
the MSTB system to the case of $N$-interacting scalar fields.
Non-linear waves in relativistic field theories with $N\geq 3$
scalar fields were sketchily described for the first time in
Reference \cite{11}. In this work, we offer a detailed analysis of
this issue. The following points merit emphasis:

\begin{itemize}

\item a) The dynamical system that encodes the solitary waves of the model
as separatrix trajectories has $N$ first integrals in involution
and hence is completely integrable. Passing from Cartesian to
Jacobi elliptic coordinates in the \lq\lq internal" space, ${\bf
R}^N$, the dynamical system becomes Hamilton-Jacobi separable. All
the kink trajectories, and hence all the solitary waves, are then
found by a special choice of the separation constants.

\item b) Deep insight into the structure of the kink manifold is gained by focusing
on the $N=3$ case. There are three kinds of kinks: 1. A
two-parameter family of topological kinks with three non-null
components that are \lq\lq generic", i.e. they are not fixed under
the action of the ${\bf Z}_2\times {\bf Z}_2\times {\bf Z}_2$
group generated by $\phi_a \to (-1)^{\delta_{ab}}\phi_a$, for
$b=1,2,3$. 2. Four one-parameter families of \lq\lq enveloping"
non-topological kinks, also with three non-null components. The
four families are related through the action of one ${\bf
Z}_2\times {\bf Z}_2$ sub-group and, together, form the envelop of
the separatrix trajectories. 3. All the solitary waves of the
$N=2$ MSTB model appear \lq\lq embedded" twice; once in each plane
containing the two ground states. Different ${\bf Z}_2$ sub-groups
leave
 these embedded kinks invariant.

\item c) The structure of the kink manifold of the $O(N)$ system with both
explicit and spontaneous symmetry breaking repeats the patterns
shown in the $N=2$ (MSTB) model and its generalization for $N=3$.
There are also generic, enveloping and embedded kinks, although
when $N$ increases the complexity of the kink manifold also
increases. For instance, the $N-1$ kink manifold is embedded $N-1$
times in the manifold of kinks of the deformed linear $O(N)$-sigma
model.

\item d) In a remarkable system obtained from the generalized  MSTB model
by also allowing asymmetries in the quartic terms of the
potential, only the embedded and enveloping topological kinks
living on singular edges survive as solitary wave solutions. In
this system, proposed in Reference \cite{24} for the $N=2$ case,
the energy of all the above topological kinks is exactly the same.
Together with vacuum degeneration, there is therefore kink
degeneration, a phenomenon that deserves further analysis.

\end{itemize}

\medskip

\noindent {\bf C}.

Solitary waves of the kind that we are to describe play an
important r\^ole in condensed matter physics. Phase transitions
characterized through order parameters of the vector type are
understood in terms of the linear (or non linear) $O(N)$-sigma
model. The order parameter is organized in the fundamental
representation of $O(N)$ and the system becomes non-linear when
this $N$-vector is forced to take its values in the coset space
${\cal M}=O(N)/O(N-1)$. In $(1+1)$-dimensional space-time, kinks
are accompanied by the fermion fractionization phenomenon
\cite{15}; this describes the continuous approximation to the
bizarre behaviour of certain one-dimensional polymers such as
poly-acetilene. When the spatial dimension is 3, as in the real
world, kinks become domain walls which are thus related to
theories involving spontaneous breaking of discrete symmetries.
This happens in the hot Big Bang cosmology, where domain wall
topological defects can be formed in a phase transition occurring
in the expansion of the very early Universe; see \cite{vs}. More
recently, domain walls have been characterized as BPS states of
SUSY gluodynamics and the Wess-Zumino model, \cite{ds}. In all
these cases there are sets of scalar fields, as in our system,
that presents a variety of domain walls with different
characteristics when seen from a 3-dimensional perspective.

In quantum field theory, the linear $O(N)$-sigma model describes
systems with spontaneous symmetry breakdown to an $O(N-1)$
sub-group and $N-1$ Goldstone bosons in the particle spectrum. At
the beginning of the sixties Gell-Mann and L\`evy analyzed low
energy hadronic phenomenology by introducing an effective
Lagrangian field theory of this type \cite{16}. Besides becoming
the central element of current algebra, linear sigma models also
enter fundamental physics in the Higgs sector of gauge theories
for elementary particle physics, see report \cite{17} for a
comprehensive review (of the perturbative sub-sector). For
instance, the linear $O(4)$-sigma model corresponds to the Higgs
sector of the electro-weak theory, while the $O(24)\times O(5)$
case provides the bosonic sector of the $SU(5)$ Grand Unified
Theory.

Either considered on their own or forming part of Gauge theories,
there are reasons to discuss deformations of the linear sigma
model. In the phenomenological approach, pions are identified with
the Goldstone bosons of the model; a deformation is then necessary
to convert these massless excitations in pseudo-Goldstone
particles, accounting for the pions light mass. Gauge theories are
today found in the low energy limit of (fundamental) string
theory. Even though deformations in the bosonic sector of gauge
theories produced by small mass terms spoil renormalizability, the
low energy features remain (almost) untouched and it is (almost)
legitimate to trust them.

Here, we shall search for domain walls when these mild
deformations are performed in the linear $O(N)$-sigma model. It is
precisely in this kind of model where the cosmological problem of
wall domination is avoided \cite{18}. Moreover, the system has a
rich manifold of topological and non-topological solitons,
allowing for topological defects with \lq\lq internal" structure
and leading to the existence of defects inside defects, a
situation that generalizes a proposal of Morris \cite{19}.

The organization of the paper is as follows: In Section \S 2 we
discuss the particle spectrum of the deformed linear $O(N)$-sigma
model as well as the manifold of the solitary wave solutions of
the system. Section \S 3 is devoted to the $N=3$ case, which is
described in full detail. We describe the situation of the
generalized MSTB model for any $N$ in Section \S 4 and briefly
discuss the phenomenon of kink degeneration in the Bazeia system.
Finally, some conclusions are drawn and some new prospects opened
in Section \S 5. An appendix on elliptic Jacobi coordinates is
also offered.

\section{Kinks in the deformed linear $O(N)$-sigma model}

In a generic sense we understand  \lq\lq kinks" as the solitary
waves of a relativistic (1+1)-dimensional scalar field theory. We
shall stick to the standard definition of solitary waves; see
\cite{Ra} and \cite{6}:

A solitary wave is a non-singular solution of the non-linear
coupled field equations of finite energy such that their energy
density has a space-time dependence of the form:
\[
\varepsilon(\vec{x},t)=\varepsilon(\vec{x}-\vec{v} t)
\]
where $\vec{v}$ is some velocity vector.

Given one $N$-component scalar field, which is a map from the
${\bf R}^{1,1}$ Minkowski space-time to ${\bf R}^N$,
$\vec{\chi}(x,t)\equiv (\chi_1(x,t),$
$\chi_2(x,t),\dots,\chi_N(x,t))$, the dynamics of the system is
governed by the action:
\[
S=\int d^2 y \left\{ \frac{1}{2}
\partial_{\mu}\vec{\chi}\cdot \partial^{\mu}\vec{\chi}-\bar{V}
(\vec{\chi})\right\}
\]

Here, $\mu=0,1$ are indices in the space-time and we shall use
$a=1,2,\dots, N$ to label components of the field in the \lq\lq
internal" ${\bf R}^N$ space in such a way that $\vec{\chi}\cdot \vec{\chi}=\displaystyle \sum_{a=1}^N \chi_a\chi_a$. In ${\bf R}^{1,1}$ we choose the
metric as $g=\left( \begin{array}{cc} 1&0\\
0&-1\end{array}\right)$ and the Einstein convention will be used
throughout the paper only for the indices in ${\bf R}^{1,1}$. The potential energy
density is:
\[
\bar{V}(\chi_1,\dots,\chi_N)=\frac{\lambda^2}{4} \left( \vec{\chi}\cdot
\vec{\chi}-\frac{m^2}{\lambda^2}\right)^2 +\sum_{a=1}^N \frac{\beta_a^2}{4}
\chi_a^2
\]
where $\lambda,m$ and $\beta_a$ are coupling constants of inverse
length. The linear $O(N)$-sigma model corresponds to the case
$\beta_a=0$, $\forall a$, which exhibits maximum $O(N)$ symmetry.
We shall focus on the deformation of this system, which is
maximally non-isotropic in the harmonic terms,  i.e. $\beta_a\neq
\beta_b$, $\forall a\neq b$. Somehow, the deformation is natural
from a quantum field theoretical vantage point as we shall explain
later and, moreover, we shall stick to the range $\beta_a^2<m^2$,
$\forall a$, in the parameter space because in this regime the
structure of the kink manifold is richer.

Introducing non-dimensional variables $\chi\to
\frac{m}{\lambda}\phi$, $y_\mu\to \frac{\sqrt{2}}{m} x_\mu$ and
$\frac{\beta_a^2}{m^2}\to \sigma_a^2$, we find our expression for
the action to be:
\begin{eqnarray}
&& S=\frac{m^2}{\lambda^2} \int d^2 x \left\{ \frac{1}{2}
\partial_\mu \vec{\phi}\cdot \partial^\mu\vec{\phi}-V(\phi_1,\dots,\phi_N)\right\}
\nonumber \\ && \label{1} \\ && V(\phi_1,\dots,\phi_N)=\frac{1}{2}
\left( \vec{\phi}\cdot \vec{\phi}-1\right)^2 +\sum_{a=1}^N\frac{1}{2}
\sigma_a^2\phi_a^2\nonumber
\end{eqnarray}

\subsection{Configuration space and particle spectrum}

The Cauchy problem for the field equations
\begin{equation}
\Box \phi_a=-\frac{\partial V}{\partial \phi_a},\quad
a=1,2,\dots,N \label{5}
\end{equation}
is fixed by choosing a \lq\lq point" $ \vec{\phi}(x,t_0)\in {\rm
Maps} ({\bf R},{\bf R}^N)$ in the configuration space
 ${\cal C}$ , and its \lq\lq tangent", $\dot{\vec{\phi}}(x,t_0)\in
T_{\vec{\phi}}{\rm Maps}({\bf R},{\bf R}^N)$, as initial
conditions to solve the system (\ref{5}) of non-linear PDE .

The configuration space itself is isomorphic to the space of
finite energy static configurations; if
$\vec{\phi}(x,t)=\vec{q}(x)$, ${\cal C}$ is the set of continuous
maps $\vec{q}:{\bf R}\to {\bf R}^N$ ($\vec{q}\equiv
(q_1,\dots,q_N)$) such that the (static) energy is finite:
\begin{equation}
E=\frac{m^3}{\lambda^2\sqrt{2}} \int dx \left\{ \frac{1}{2}
\frac{d\vec{q}}{dx}\cdot
\frac{d\vec{q}}{dx}+V(\vec{q})\right\}<+\infty ;\label{2}
\end{equation}
thus, ${\cal C}=\left\{ \vec{q}(x)/E[\vec{q}]<+\infty\right\} $.
$\vec{q}(x)\in{\cal C}$ only if $\vec{q}$ satisfies the asymptotic
conditions:
\begin{equation}
\lim_{x\to \pm \infty}\frac{dq_a}{dx}=0\  ,\qquad \lim_{x\to \pm
\infty}q_a(x)=v_a,\  \forall a=1,\dots,N\label{3}
\end{equation}
where $\vec{v}\equiv (v_1,\dots,v_N)$ is a constant vector that
belongs to the set ${\cal M}$ of vectors annihilating $V$. We
assume, without loss of generality, the following ordering in the
space of parameters: $\sigma_1=0<\sigma_2<\dots <\sigma_N<1$.
${\cal M}$ is thus formed by two vectors
\begin{equation}
{\cal M}=\left\{ \vec{v}^{\, \pm}=(\pm 1,0,\dots,0)\right\}\label{4}
\end{equation}
which are the absolute minima of $V$.

We refer to ${\cal M}$ as the vacuum manifold because in the
quantum version of the theory points in ${\cal M}$ are the
expectation values of the quantum field operators $\hat{\phi}_a$
at the ground states (\lq\lq vacua") of the system. The vacuum
degeneration - i.e.the existence of more than one vector in ${\cal
M}$ - is related to the breaking of symmetry. Besides
two-dimensional Poincar\'e invariance, there is a \lq\lq internal"
symmetry with respect to the discrete group $G={\bf Z}_2\times
\stackrel{N}{\dots} \times{\bf Z}_2={\bf Z}_2^{\times N}$
generated by $\phi_a \to (-1)^{\delta_{ab}}\phi_a$, for
$b=1,2,\dots,N$, $\forall a=1,\dots,N$. The vacuum manifold is the
orbit of one element by the group action
\[
{\cal M}=G/H_{\vec{v}^\pm}={\bf Z}_2,\qquad
H_{\vec{v}^\pm}\vec{v}^\pm=\vec{v}^\pm
\]
$H_{\vec{v}^\pm}={\bf Z}_2\times \stackrel{N-1}{\dots}\times {\bf
Z}_2$ is the little group of the vacuum $\vec{v}^{\pm}$. The
generators of $H_{\vec{v}^\pm}$ are the transformations $\phi_a
\to (-1)^{\delta_{ab}}\phi_a$, for $b=1,2,\dots,N$, and
$a=2,3,\dots,N$, so that $H_{\vec{v}^\pm}$ survives as a symmetry
group when quantizing around $\vec{v}^\pm$. We can understand the
internal parity group $G$ as the discrete \lq\lq gauge" symmetry:
in (1+1)-dimensions no dynamical degrees of freedom related to
gauge potentials appear.

Vectors in ${\cal M}$ are critical points of $V$ satisfying
$\left.\frac{\partial V}{\partial
\phi_a}\right|_{\vec{\phi}=\vec{v}^\pm}=0$ and therefore constant
solutions of the field equations (\ref{5}).
The plane wave
expansion around $\vec{\phi}^{\pm}(x,t)=\vec{v}^{\pm}$
\[
\phi_{v_a^\pm}(x,t)=v_a^\pm +\sum_k A_a^\pm (k) e^{i\omega t-ikx}
\]
is a solution of (\ref{5}) if the dispersion relation
\begin{equation}
\delta_{ab}\omega^2=\delta_{ab} k^2+M_{ab}^2(\vec{v}^{\pm})\
,\quad M_{ab}^2=\frac{\partial^2 V}{\partial \phi_a
\partial\phi_b}(\vec{v}^\pm)\label{6}
\end{equation}
holds.

In the quantum theory, these plane waves become the fundamental
quanta with mass matrix $M_{ab}^2(\vec{v}^\pm)$ and one reads the
particle spectrum at a chosen critical point of $V$ from
(\ref{6}). Because $\vec{v}^\pm\in {\cal M}$ are minima of $V$
there are no negative eigenvalues of $M_{ab}^2(\vec{v}^\pm)$ and
the dependence on time of the plane waves around $\vec{v}^\pm$ is
bounded: $e^{i\omega t}$. The choice of $\vec{v}^\pm$ as the
starting point of the quantization procedure \lq\lq spontaneously"
breaks the symmetry $G={\bf Z}_2^{\times N}$ of the action to
$H_{\vec{v}^\pm}={\bf Z}_2^{\times (N-1)}$, which is the remaining
one that survives in the particle spectrum.

In our model, we read the particle spectrum from:
\begin{equation}
M^2(\vec{v}^\pm)=\frac{m^2}{2} \cdot \left( \begin{array}{cccc} 4
& 0 & \dots & 0\\ 0 & \sigma_2^2 & \dots & 0\\ \vdots & \vdots &
\ddots &\vdots \\ 0 & 0& \dots & \sigma_N^2 \end{array} \right)
\label{7}
\end{equation}

Considering this system as a physical description of the continuum
approximation to a one-dimensional crystal with an $N$-component
order parameter, the particle spectrum describes a single phase
with $N$ phonon branches. We see explicitly how the symmetry group
$G={\bf Z}_2^{\times N}$ is \lq\lq broken" by the choice of the
$\vec{v}^\pm$ vacuum to the $H_{\vec{v}^\pm}={\bf Z}_2^{\times
(N-1)}$ sub-group: the $N$ phonon branches have different masses
or \lq\lq energy gaps". From the point of view of particle physics
we can say that there are no tachyons; only a pseudo-Goldstone
particle becomes a Goldstone boson if the corresponding $\sigma_a$
goes to zero.

It is interesting to see the model as a member of the family
characterized by the potential energy densities:
\[
V=\frac{1}{2} \left( \sum_{a,b=1}^N \alpha_{ab} \phi_a \phi_b -\gamma^2 \right)^2
+\sum_{a,b=1}^N \frac{\sigma_{ab}}{2} \phi_a \phi_b
\]
where $\alpha_{ab}, \sigma_{ab}$ and $\gamma^2$ are \lq\lq bare"
non-dimensional parameters. Ultraviolet divergences are controlled
by normal ordering in the quantum theory, but the need arises to
introduce a renormalization \lq point\rq\,  $\mu^2$, and the
dependence of the renormalized parameters on $\mu^2$ is determined
by the renormalization group equation. One special solution, a
specific renormalization group flow, might lead to the \lq\lq
point":
\[
\alpha_{ab}^R(\mu^2) =\delta_{ab};
\quad \sigma_{ab}^R(\mu^2)=0;\quad \gamma^R(\mu^2)=1
\]
in the space of quantum field theory models in the family. This
point is the linear $O(N)$-sigma model which has $G=O(N)$ as the
(continuous) symmetry group. The vacuum orbit is, however, ${\cal
M}=O(N)/O(N-1)=S^{N-1}$, the $(N-1)$-dimensional sphere, and thus
there is no unbroken symmetry left: there are $N-1$ massless
particles. If the only modification of the renormalized parameters
is to allow for non-zero values of $\sigma_{ab}^R(\mu^2),\
a=b=r+1,\dots,N$ there are still $r-1$ Goldstone bosons.

Coleman \cite{9} established that in $(1+1)$-dimensions the
infrared asymptotics of the two-point Green functions of a quantum
scalar field forbids poles at $\omega^2=k^2$; there are no
Goldstone bosons in $(1+1)$-dimensions. It is thus impossible to
reach the $O(N)$-sigma model or its deformation with the $O(r)$
symmetry spontaneously broken to $O(r-1)$ in the renormalization
group flow. The closest admissible points are the models
characterized by:
\[
\alpha_{ab}^R(\mu^2)=\delta_{ab},\quad
\gamma^R(\mu^2)=1,\sigma_{ab}^R=0,a\neq b
\]
\[
\sigma_{11}^R(\mu^2)=0<\sigma_{22}^R(\mu^2)=\sigma_2^2\leq \dots
\leq \sigma_{NN}^R(\mu^2)=\sigma_N^2<1
\]

In this paper we shall focus on the case of maximal explicit
symmetry breaking; i.e. when strict inequalities in the parameter
space occur. Nevertheless, we shall comment on the allowed
situation characterized by
\[
\sigma_1^2=0<\sigma_2^2=\dots=\sigma_{r_1}^2<\sigma_{r_1+1}^2=\dots
=\sigma_{r_2}^2<\dots <\sigma_{r_k+1}^2=\dots =\sigma_N^2<1
\]
when there is degeneration in the spectrum but no Goldstone
bosons. Note that the generators of the $O(r_1-1)\times
O(r_2-r_1)\times \dots \times O(N-r_k)$ symmetry sub-group are in
the little group of the vacuum. The symmetry group is $G={\bf
Z}_2\times O(r_1-1)\times O(r_2-r_1)\times \dots \times O(N-r_k)$,
$H_{\vec{v}^\pm}= O(r_1-1)\times O(r_2-r_1)\times \dots \times
O(N-r_k)$ and the vacuum orbit is ${\cal M}=G/H_{\vec{v}^\pm}={\bf
Z}_2$.

\subsection{Configuration space topology: kinks and dynamical systems}

The configuration space of the model is the union of topologically
disconnected sectors: ${\cal C}={\displaystyle
\bigsqcup_{\alpha,\beta=1}^2} {\cal C}^{\alpha\beta}$; thus,
$\pi_0({\cal C})={\bf Z}_2\times {\bf Z}_2$ and  $|\pi_0({\cal
C})|=4$ are  respectively the zeroth-order homotopy group of
${\cal C}$ and its order. This comes from the asymptotic
conditions (\ref{3}) and the continuity of the time evolution .
There are topological charges defined for each configuration in
${\cal C}$ as:
\[
Q_a^T=\frac{1}{2} \int_{-\infty}^{\infty} dx
\frac{d\phi_a}{dx}=\frac{1}{2} \left(
\phi_a(+\infty,t)-\phi_a(-\infty,t)\right)
\]

It should be noted that  $Q_a^T$ is independent of $t$, $\forall
a$, and in our system equal to zero if $a\geq 2$. Therefore the
four sectors ${\cal C}^{\alpha\beta}$ are labelled by the values
$\alpha, \beta$ of the fields at infinity compatible with finite
energy and $ Q_1^T$ determines the homotopy class in $\pi_0({\cal
C})={\bf Z}_2\times {\bf Z}_2$.

The critical points of $E$ are time-independent finite-energy
solutions of the field equations. If they are not spatially
homogeneous, the critical points correspond to solitary waves that
are therefore related to the topological structure of ${\cal
C}^{\alpha \beta}$. Besides complying with (\ref{3}), solitary
waves satisfy the system of ordinary differential equations:
\begin{equation}
\frac{d^2q_a}{dx^2}=\frac{\partial V}{\partial q_a}\label{8}
\end{equation}
Recall that $\phi_a(x,t)=q_a(x)$. Solving the system (\ref{8}) is
tantamount to finding the solutions of the Lagrangian dynamical
system in which $x=\tau$ plays  the r\^ole of time, the \lq\lq
particle" position is determined by $q_a(\tau)$, and the potential
energy of the particle is $U(\vec{q})=-V(\vec{q})$. From this
perspective the static field energy $E$ is seen as the particle
action:
\begin{equation}
E=J=\int d\tau \left\{ \frac{1}{2} \frac{d\vec{q}}{d\tau}\cdot
\frac{d\vec{q}}{d\tau}-U(\vec{q})\right\} \label{9}
\end{equation}

Trajectories that behaves asymptotically in the $\tau$-time as
ruled by (\ref{3}) have a finite action, $J$, in the mechanical
problem and are in one-to-one correspondence with solitary
waves/kinks that have energy $E=J$ in the field theoretical
system.

The mechanical analogy is very helpful when one is dealing with a
real scalar field theory because, then, a first integral is all
that we need to find all the solutions. Vector scalar fields of
$N$ components lead to $N$-dimensional dynamical systems which are
seldom solvable. Magyari and Thomas \cite{4} realized that the
two-dimensional dynamical system arising in connection with the
MSTB model is a completely integrable one in the Liouville sense;
there are two first integrals in involution. Moreover, Ito
\cite{5} has shown that the mechanical system is Hamilton-Jacobi
separable, finding all the trajectories and hence all the kinks of
the MSTB model. In a recent publication \cite{6}, we have
developed this procedure for two $N=2$ models with interesting
features: the first system is a deformation of the
(1+1)-dimensional scalar field theory, where the potential energy
density is the Chern-Simons-Higgs potential arising in self-dual
planar gauge theories. The second one is a deformation of the
linear $O(2)$-sigma model, which is different from the MSTB model.

To extend this method of finding kinks to the linear $O(N)$-sigma
model, $N\geq 3$, deformed in such a way that the $O(N)$ symmetry
is explicitly broken to $G={\bf Z}_2^{\times N}$, we start from
the \lq\lq particle" action:
\[
J=\int d\tau \left\{ \frac{1}{2}
\frac{d\vec{q}}{d\tau}\cdot\frac{d\vec{q}}{d\tau} +\frac{1}{2}
\left(\vec{q}\cdot\vec{q} -1\right)^2+\frac{1}{2}\sum_{a=1}^N
\sigma_a^2 q_a^2 \right\}=\int d\tau {\cal
L}(\vec{q},\dot{\vec{q}})
\]

The particle motion equations are:
\begin{equation}
\frac{d^2q_a}{d\tau^2}=2 q_a (\vec{q}\cdot \vec{q}-1)+\sigma_a^2 q_a,\quad \forall a=1,\dots,N\label{10}
\end{equation}
which are mathematically identical to the field equations for
static configurations. Finite action trajectories, kinks in the
field theory, should also satisfy the asymptotic conditions:
\begin{equation}
\lim_{\tau\to \pm \infty} \frac{dq_a}{d\tau}=0,\qquad
\lim_{\tau\to \pm \infty}q_a(\tau)=\pm \delta_{a1}\label{3b}
\end{equation}

We shall use the Hamiltonian formalism to integrate the mechanical
system. The canonical momenta $p_a(\tau)=\frac{\partial {\cal
L}}{\partial \dot{q}_a}=\frac{dq_a}{d\tau}(\tau)$, together with
the positions $q_a(\tau)$, form a system of local coordinates in
phase space. We should bear in mind that
$p_a(\tau)=\frac{d\phi_a}{dx}$ when going back to the field
theory. The mechanical Hamiltonian
\begin{equation}
I_1=\frac{1}{2} \vec{p}\cdot \vec{p}-\frac{1}{2} \left( \vec{q}\cdot \vec{q}-1\right)^2-\sum_{a=1}^N \frac{1}{2} \sigma_a^2 q_a^2\label{11}
\end{equation}
leads to the system of canonical equations
\[
\frac{dq_a}{d\tau}=\{ I_1,q_a\},\qquad \frac{dp_a}{d\tau}=\{
I_1,p_a\}
\]
equivalent to (\ref{10}). Given any two functions
$F(\vec{q},\vec{p})$, $G(\vec{q},\vec{p})$ in phase space, the
Poisson bracket is defined in the usual way:
\[
\{ F,G\}=\sum_{a=1}^N\left( \frac{\partial F}{\partial q_a}\cdot
\frac{\partial G}{\partial p_a}-\frac{\partial F}{\partial
p_a}\cdot \frac{\partial G}{\partial p_a}\right)
\]

Obviously $\frac{dI_1}{d\tau}=0$, but our mechanical system is
full of other invariants. In fact,  as early as 1919 Garnier
\cite{21} solved the motion equations and described periodic
trajectories in terms of Theta functions: the kink trajectories of
finite \lq\lq action" correspond
 to a limiting case and are the
separatrices between the periodic trajectories and unbounded
motion. More recently Grosse, and other authors \cite{22} have
shown that the functions:
\begin{equation}
K_a=\sum_{b=1,b\neq a}^N \frac{1}{\sigma_b^2-\sigma_a^2}
l_{ab}^2+p_a^2+(2-\sigma_a^2) q_a^2-q_a^2\sum_{b=1}^N
q_b^2\label{12}
\end{equation}
\[
l_{ab}=p_aq_b-p_bq_a
\]
are first integrals in involution:
\[
\{ I_1,K_a\}=0\qquad \{ K_a,K_b\}=0
\]

There is a set of $N+1$ invariants in involution:
$I_1,K_1,K_2,\dots,K_N$. The dynamical system is not
superintegrable, however, because there are only $N$-independent
invariants: $K_1+K_2+\dots+K_N=2I_1+1$. According to the Liouville
theorem, the $N$-dimensional mechanical system is completely
integrable and all trajectories can be found, at least in
principle.

At this point we pause to explain the singular nature of the
deformation of the linear $O(N)$-sigma model chosen from among
many possibilities. The $\frac{\sigma_a^2}{2}\phi_a^2$ terms
explicitly break the $O(N)$-symmetry of the linear sigma model;
the case $\sigma_a^2=0$, $\forall a=1,2,\dots,N$. In the
mechanical system the $O(N)$ internal transformations become
ordinary rotations. The angular momentum components, $l_{ab}$,
conserved in the limit $\sigma_a^2=0$, $\forall a$, are no longer
\lq time\rq-independent if $\sigma_a^2\neq 0$. There are, however,
$N$ invariants $K_a$, which in the limit $\sigma_a=0$, $\forall
a$, are given in terms of the $O(N)$-invariants: the $r$ Casimir
invariants and the $r$ generators of the Cartan sub-algebra, where
$r=\frac{N}{2}$ or $\frac{N-1}{2}$ if $N$-even or -odd is the rank
of the group. A warning: in the $N=$ odd case, the energy must be
added to the other $N-1$ invariants built from the Cartan
sub-algebra and the Casimir invariants. For any $N$, the maximally
asymmetric chosen deformation is special because it retains enough
symmetry to solve the mechanical system. There is no Lie algebra
associated with $K_a$ however; since the invariants are quadratic
in $q_a$, $p_a$, the action of $K_a$ in the phase space, given by
$\{ K_a,q_b\}$ and $\{ K_a,p_b\}$, is non-linear.

In (1+1)-dimensional field theory, the energy-momentum tensor:
\[
T^{\mu\nu}=\frac{\partial {\cal L}}{\partial
(\partial_\mu\phi_a)}\cdot \partial^{\nu}\phi_a-g^{\mu\nu}{\cal L}
\]
is divergenceless due to invariance under space-time translations.
$P^{\mu}=\int dx T^{0\mu}$ are thus conserved quantities whatever
the values of $\sigma_a^2$. The $O(N)$ \lq\lq isospin" currents
however,
\[
J_a^{\mu}=\sum_{b,c=1}^N c_{abc}\phi_b \partial^{\mu}\phi_c
\]
are only divergenceless if $\sigma_a=0$, $\forall a$. The
$c_{abc}$ are the Lie $O(N)$ structure constants and the charges
$Q_a=\int dx J_a^0$ are not conserved if there is no symmetry with
respect to the transformation generated by them. For static
configurations, we have
\[
\int dx T^{00}=E=J,\quad T^{10}=T^{01}=0,\qquad T^{11}=I_1
\]
\[
J_a^0=0,\qquad J_a^1=\sum_{b,c=1}^N c_{abc} l_{bc}
\]

In terms of the \lq isospin\rq  currents, the invariants $K_a$ can
be written as:
\[
K_a=\sum_{b=1,b\neq a}^N \frac{1}{\sigma_b^2-\sigma_a^2} \cdot
\left( \sum_{c=1}^N c_{abc}J_c^1\right) \, \left( \sum_{d=1}^N c_{abd} J_d^1\right) +\left( \frac{\partial \phi_a}{\partial x}\right)^2 +(2-\sigma_a^2) \phi_a^2
-\phi_a^2 \sum_{b=1}^N \phi_b^2
\]

We expect that the time-evolution occurs in such a way that there
is some equation of non-linear character
\[
F\left( \frac{\partial L_a}{\partial t},\frac{\partial
K_a}{\partial x}\right)=0
\]
between $K_a$ and
\[
L_a=\sum_{b=1,b\neq a}^N \frac{1}{\sigma_b^2-\sigma_a^2}\cdot
\left( \sum_{c=1}^N c_{abc}J_c^0\right) \, \left( \sum_{d=1}^N c_{abd}J_d^0\right) +\left( \frac{\partial \phi_a}{\partial t}\right)^2 +(2-\sigma_a^2)
\phi_a^2-\phi_a^2 \sum_{b=1}^N \phi_b^2
\]
which reduces to $\frac{\partial J_a^0}{\partial t}=\frac{\partial
J_a^1}{\partial x}$ when $\sigma_a=0$, $\forall a$. The situation
is analogous to that occurring between conformal field theories
and models with infinite-dimensional algebraic symmetry as in
(1+1)-dimensional Toda field theories and Toda affine models
\cite{23}. There are two differences: (1) the conformal group is
infinite dimensional in (1+1)-dimensions. We have only one
finite-dimensional group $O(N)$ and thus we can solve only the
static limit of the field theory model. (2) Due to the non-linear
character of the deformation of the $O(N)$ Lie generators, we do
not even have a finite-dimensional Lie algebra.

\subsection{The Hamilton-Jacobi equation and kink trajectories}

The $K_a$ invariants defined in (\ref{12}) are quadratic in the
momenta, but not orthogonal (they contain terms in $p_ap_b, a\neq
b$). Therefore, the St\"ackel theorem can not be applied to assure
Hamilton-Jacobi separability. This problem is surpassed in our
dynamical system with the choice of some suitable system of
coordinates. The appropriate system is provided by elliptic Jacobi
coordinates, with a choice of separation constants determined by
the deformation parameters giving mass to the Goldstone bosons;
$\bar{\sigma}_a^2=1-\sigma_a^2\  ,\quad \forall a=1,2,\dots,N$.
Thus we define:
\begin{equation}
q_{a}^2=\frac{\displaystyle \prod_{b=1}^N
(\bar{\sigma}_a^2-\lambda_b)}{\displaystyle \prod_{b=1,b\neq a}^N
(\bar{\sigma}_a^2-\bar{\sigma}_b^2)}=
\frac{\Lambda(\bar{\sigma}_a^2)}{A'(\bar{\sigma}_a^2)}\label{13}
\end{equation}
ruling the change of coordinates from Cartesian, $\vec{q}\equiv
(q_1,\dots,q_N)$ to elliptic $\vec{\lambda}\equiv
(\lambda_1,\dots,\lambda_N)$. In the Appendix, it is explained how
the elliptic variables are split:
\begin{equation}
-\infty<\lambda_1<\bar{\sigma}_N^2<\lambda_2<\bar{\sigma}_{N-1}^2<\dots
<\bar{\sigma}_2^2<\lambda_N<1\label{14}
\end{equation}
Notice that formula (\ref{13}) coincides with formula (\ref{A5})
in the Appendix if we change $q_a$ by $q_{N-a+1}$ and choose
$r_{N-a+1}=\bar{\sigma}_a^2$.

Together with formula (\ref{13}), this splitting means that the
change of coordinates produces a map from a sub-space of ${\bf
R}^N$, characterized as the set of points which are not invariants
under the ${\bf Z}_2^{\times N}$ group generated by $q_a\to
(-1)^{\delta_{ab}}q_a, b=1,\cdots,N$, to the interior of the
infinite parallelepiped $P_N(\infty)$ obtained by replacing the
inequalities in (\ref{14}) by equalities: $-\infty<
\lambda_1\leq\bar{\sigma}_N^2\leq\dots
\leq\bar{\sigma}_2^2\leq\lambda_N\leq1$. Notice that in this map
$2^N$ regular points in ${\bf R}^N$ go to a single point in the
interior of $P_N(\infty)$; Singular points lie in the ${\bf R}^m$,
$m=0,1,\dots,N-1$, sub-spaces that are invariant under the action
of some non-trivial element of $G={\bf Z}^{\times N}$. These
singular sub-spaces are mapped into the boundary of $P_N(\infty)$.

The standard length of an interval in Euclidean space is expressed
in elliptic coordinates in the form
\[
ds^2=\sum_{a=1}^N dq_a dq_a=\sum_{a=1}^N g_{aa}(\vec{\lambda})
d\lambda_a d\lambda_a
\]
because the metric $g_{aa}(\vec{\lambda})$, as derived in the
Appendix, is:
\[
g_{aa}(\vec{\lambda})=-\frac{1}{4}
\frac{f_a(\vec{\lambda})}{A(\lambda_a)},\quad
g_{ab}(\vec{\lambda})=0,\forall a\neq b
\]
where $\displaystyle
A(\lambda_a)=\prod_{b=1}^N(\lambda_a-\bar{\sigma}_b^2)$, and
\footnote{The standard notation in the literature on elliptic
Jacobi coordinates is
$\displaystyle\Lambda'(\lambda_a)=\prod_{b=1\atop b\neq
a}^N(\lambda_a-\lambda_b)$, see Appendix. We shall use
$f_a(\vec{\lambda})$ in the main text instead of
$\Lambda'(\lambda_a)$, to stress the fact that this quantity
depends on all the components of $\vec{\lambda}$}
\[
f_a(\vec{\lambda})=f_a(\lambda_1,\lambda_2,\cdots,\lambda_N)=\prod_{b=1\atop
b\neq a}^N(\lambda_a-\lambda_b)
\]

Therefore, the Lagrangian reads:
\begin{eqnarray}
L&=& \frac{1}{2} \sum_{a=1}^N \left(
\frac{dq_a}{d\tau}\right)^2-U(q_1,\dots,q_N)\nonumber\\
&=&\frac{1}{2} \sum_{a=1}^N g_{aa}(\vec{\lambda}) \left(
\frac{d\lambda_a}{d\tau}\right)^2-U_{\lambda}(\lambda_1,\dots,\lambda_N)\label{15}
\end{eqnarray}
where the potential in elliptic coordinates is:
\begin{equation}
U_{\lambda}(\lambda_1,\dots,\lambda_N)=-\sum_{a=1}^N \frac{1}{2}
\frac{\lambda_a^{N+1}-(\alpha-1)\lambda_a^N+(1-\alpha+\beta)\lambda_a^{N-1}}{f_a(\vec{\lambda})}\label{16}
\end{equation}
\[
\alpha=\sum_{a=1}^N \bar{\sigma}_a^2\  ,\qquad
\beta=\sum_{a=1,a<b}^N\sum_{b=2}^N \bar{\sigma}_a^2
\bar{\sigma}_b^2
\]

The computation of $U_{\lambda}(\vec{\lambda})$ is highly
non-trivial and requires the use of formulas that follow the
Jacobi Lemma, such as (\ref{A8b}), (\ref{A9}), (\ref{A10}), etc.

The canonical momenta associated to the $\lambda_a$ variables are:
\[
\pi_a=\sum_{b=1}^N g_{ab}(\vec{\lambda})
\frac{d\lambda_b}{d\tau}=g_{aa}(\vec{\lambda})
\frac{d\lambda_a}{d\tau}
\]
and, through the standard Legendre transformation, we write the
Hamiltonian:
\begin{equation}
H=\frac{1}{2} \sum_{a=1}^N \frac{-4
A(\lambda_a)}{f_a(\vec{\lambda})} \, \pi_a^2+U_\lambda
(\vec{\lambda})\label{17}
\end{equation}

The key point is that $H$ can be written in St\"ackel's form:
\begin{eqnarray}
H&=&\sum_{a=1}^N \frac{H_a}{f_a(\vec{\lambda}
)}=\label{18}\\
&=&\sum_{a=1}^N \frac{\left[ -2 A(\lambda_a)\pi_a^2-\frac{1}{2}
\left(
\lambda_a^{N+1}-(\alpha-1)\lambda_a^N+(1-\alpha+\beta)\lambda_a^{N-1}\right)
\right]}{f_a(\vec{\lambda})}\nonumber
\end{eqnarray}
such that the Hamilton-Jacobi equation
\begin{equation}
\frac{\partial {\cal S}}{\partial \tau}+H\left( \frac{\partial
{\cal S}}{\partial \lambda_1},\dots,\frac{\partial {\cal
S}}{\partial \lambda_N},\lambda_1,\dots,\lambda_N\right)
=0\label{19}
\end{equation}
is completely separable. We now prove this last statement.

Fixing $H=I_1$, the first integral of energy, and having in mind
the expression (\ref{18}) of $H$, we write the solution of
(\ref{19}) as:
\begin{equation}
{\cal S}=-I_1 \tau+\sum_{a=1}^N S_a (\lambda_a) .\label{20}
\end{equation}
Therefore, (\ref{19}) reduces to :
\begin{equation}
I_1=H\left(
\frac{dS_1}{d\lambda_1},\dots,\frac{dS_N}{d\lambda_N},\lambda_1,\dots,\lambda_N\right)=
\sum_{a=1}^N\frac{H_a}{f_a(\vec{\lambda})} \label{r1}
\end{equation}

The Hamilton-Jacobi PDE equation (\ref{19}) becomes equivalent to
the system of non-coupled ordinary differential equations
\begin{equation}
H_a\left( \frac{dS_a}{d\lambda_a},\lambda_a \right)=\eta_1
\lambda_a^{N-1}+\eta_2\lambda_a^{N-2}+\dots+ \eta_{N-1} \lambda_a
+\eta_N\label{22}
\end{equation}
where
\begin{equation}
H_a\left( \frac{dS_a}{d\lambda_a},\lambda_a \right) = -2
A(\lambda_a)\left( \frac{dS_a}{d\lambda_a}\right)^2- \frac{1}{2}
\left( \lambda_a^{N+1}
-(\alpha-1)\lambda_a^N+(1-\alpha+\beta)\lambda_a^{N-1}\right),\label{r2}
\end{equation}
due to the identity
\begin{equation}
I_1=I_1 \sum_{a=1}^N
\frac{\lambda_a^{N-1}}{f_a(\vec{\lambda})}+\eta_2
\sum_{a=1}^N\frac{\lambda_a^{N-2}}{f_a(\vec{\lambda})}+\dots
+\eta_N \sum_{a=1}^N\frac{1}{f_a(\vec{\lambda})}\nonumber
\end{equation}
-observe that $\eta_1=I_1$- which follows from the Jacobi Lemma
and the subsequent relations (Appendix)
\[
\sum_{a=1}^N \frac{\lambda_a^{N-1}}{f_a(\vec{\lambda})}=1\ ,\qquad
\sum_{a=1}^N \frac{\lambda_a^{N-i}}{f_a(\vec{\lambda})}=0, \forall
i=2,\dots,N
\]
Alternatively, one could take a more direct approach to show
formula (\ref{22}). We start from (\ref{r1}), written explicitly
as
\begin{eqnarray}
\frac{H_1(\lambda_1)}{(\lambda_1-\lambda_2)(\lambda_1-\lambda_3)\cdots(\lambda_1-\lambda_N)}&+&
\frac{H_2(\lambda_2)}{(\lambda_2-\lambda_1)(\lambda_2-\lambda_3)\cdots(\lambda_2-\lambda_N)}+\cdots\nonumber\\
&+&\frac{H_N(\lambda_N)}{(\lambda_N-\lambda_1)(\lambda_N-\lambda_2)\cdots(\lambda_N-\lambda_{N-1})}=I_1\label{r11}
\end{eqnarray}
Here, each $H_a(\lambda_a)$ is of the form (\ref{r2}) due to the
ansatz (\ref{20}). Multiplying (\ref{r11}) by
$(\lambda_1-\lambda_2)$ and setting $\lambda_1=\lambda_2$ one sees
that $H_1(\lambda)\equiv H_2(\lambda)$, and hence by symmetry, all
$H_a(\lambda)$, $a=1,2,\cdots,N$ are identical. It suffices
therefore to find $H_1(\lambda)$. Multiplying (\ref{r11}) by
$f_1(\vec{\lambda})$ one obtains
\begin{equation}
H_1(\lambda_1)+P_2(\vec{\lambda})H_2(\lambda_2)+\cdots+P_N(\vec{\lambda})H_N(\lambda_N)=I_1(\lambda_1-\lambda_2)
(\lambda_1-\lambda_3)\cdots(\lambda_1-\lambda_N),\label{r3}
\end{equation}
where each $P_a(\vec{\lambda})$, $a=2,3,\cdots,N$, is a polynomial
of degree $N-1$ in $\lambda_1$. Differentiating (\ref{r3}) $N$
times with respect to $\lambda_1$ yields
\begin{equation}
\frac{d^N H_1(\lambda_1)}{d\lambda_1^N}=N!I_1
\end{equation}
whence it follows that $H_1(\lambda)$ is a degree $N$ polynomial
in $\lambda$ with leading coefficient $I_1$, in agreement with
equation (\ref{22}).

 The set of separation constants $\eta_i, \, i=1,2,\dots,N$
is another system of first integrals in involution. They can be
expressed in the elliptic phase space $T^*P_N(\infty)$ as
functions of $\lambda_a$ and $\pi_a=\frac{dS_a}{d\lambda_a}$ by
solving the linear system of equations (\ref{22}) in the unknown
$\eta_i$, which is a Vandermonde system. The $N-1$ roots of the
polynomial
\[
\lambda_a^{N-1}+\frac{\eta_2}{I_1}\lambda_a^{N-1}+\dots+\frac{\eta_N}{I_1}=(\lambda_a-F_2)(\lambda_a-F_3)\dots
(\lambda_a-F_N)
\]
together with the energy $F_1=I_1$ form another system of
invariants in involution. Both systems are related through the
identities $\eta_1=I_1$ and:
\[
\eta_a=(-1)^a I_1 \sum_{i_1<i_2<\dots<i_a}F_{i_1}F_{i_2}\dots
F_{i_a},\  i_\alpha=2,3,\dots,N
\]
Therefore, all the separation constants $\eta_a$ are proportional
to the \lq\lq particle" energy $I_1$.

Defining the polynomial $B(\lambda_a)$ in the form:
\[
B(\lambda_a)=\lambda_a^{N+1}-
(\alpha-1)\lambda^N+(1-\alpha+\beta+2
\eta_1)\lambda_a^{N-1}+2\eta_2\lambda_a^{N-2}+\dots+2\eta_N
\]
the solution of the differential equation (\ref{22}) is a
quadrature:
\begin{equation}
S_a(\lambda_a)=\frac{1}{2} {\rm sign}\left(
\frac{dS_a}{d\lambda_a}\right) \int \sqrt{\left|
\frac{B(\lambda_a)}{A(\lambda_a)}\right|} d\lambda_a\label{23}
\end{equation}
and the general solution of the Hamilton-Jacobi equation reads:
\begin{equation}
{\cal S}=-\eta_1 \tau+\sum_{a=1}^N \frac{1}{2} {\rm sign}\left(
\frac{dS_a}{d\lambda_a}\right) \int \sqrt{\left|
\frac{B(\lambda_a)}{A(\lambda_a)}\right|}d\lambda_a\label{24}
\end{equation}

The explicit integration of the quadratures in (\ref{23}) requires
the theory of Theta functions of genus depending on $N$. The
action of the associated trajectories is infinite because they are
either periodic or unbounded. The asymptotic conditions (\ref{3})
that guarantee finite action to continuous trajectories satisfying
them also require that the energy used by the particle in these
trajectories should be zero. This is so because $\left.
I_1\right|_{\tau=\pm \infty}=0$ and, being an invariant of the
evolution, $I_1=0,\forall \tau$.

We recall that the trajectories of finite action in an evolution
lasting an infinite time are the kinks of the parent field theory
system: one just trades the finite action of the trajectory for
finite energy of the non-linear wave. Therefore, the kinks are the
trajectories obtained when all the separation constants in
(\ref{22}) are zero: $\eta_a=0, \forall a$. These are the
separatrices between bounded and unbounded motion and the
integrals in (\ref{23}) are easier to compute.

The explicit trajectories are also provided by the Hamilton-Jacobi
principle, through the set of equations:
\[
\gamma_a=\frac{\partial {\cal S}}{\partial \eta_a}\ ,\quad
a=1,2,\dots,N
\]
where the $\gamma_a$ are integration constants.  In the
hypersurface of the phase space determined by
$\eta_1=\dots=\eta_N=0$, the first equation
\begin{equation}
\gamma_1=-\tau+\sum_{a=1}^N \frac{1}{2} {\rm sign}(\pi_a)\int
\frac{\lambda_a^{N-1}d\lambda_a}{\left|
A(\lambda_a)\right|}\sqrt{\left|
\frac{A(\lambda_a)}{\lambda_a^{N+1}
-(\alpha-1)\lambda_a^N+(1-\alpha+\beta)\lambda_a^{N-1}}\right|}\label{25}
\end{equation}
rules the time-dependence of the particle in its journey through
the orbit. From the field theoretical point of view, it provides
the kink form factor. The other $N-1$ equations, $i=2,3,\dots,N$,
\begin{equation}
\gamma_i=\sum_{a=1}^N \frac{1}{2} {\rm sign}(\pi_a)\int
\frac{\lambda_a^{N-i}d\lambda_a}{\left|
A(\lambda_a)\right|}\sqrt{\left|
\frac{A(\lambda_a)}{\lambda_a^{N+1}
-(\alpha-1)\lambda_a^N+(1-\alpha+\beta)\lambda_a^{N-1}}\right|}\label{26}
\end{equation}
determine the orbit in $P_N(\infty)$, the intersection of $N-1$
hypersurfaces in the configuration space. Therefore, there is a
$N-1$-dimensional family of kinks parametrized by the finite
values of $\gamma_i$.

Although (\ref{25}) and (\ref{26}) identify all the separatrix
trajectories of the mechanical system and henceforth all the kink
solutions of the deformed linear $O(N)$-sigma model, an explicit
description of such solitary waves is difficult for two reasons:
(1). (\ref{25}) and (\ref{26}) form a system of transcendent
equations of impossible analytical resolution. (2). Even if it
were possible, expressing back the solution in Cartesian
coordinates through (\ref{A5}) for $N\geq 3$ is another impossible
task by analytical means.

\section{$N$=3}

To gain insight into the nature of the different kinks of the
model, in this Section we shall address in full detail the $N=3$
case. We shall deal with a (1+1)-dimensional field theory
including three scalar fields which transform according to a
vector representation of the $O(3)$ group. The structure of the
solitary wave solutions of the $N=3$ system is extremely rich from
different points of view and shows the behavioural pattern of the
general case with $N$-component fields.

\subsection{The general solution of the Hamilton-Jacobi equation}

The Hamiltonian of the underlying dynamical system reads:
\begin{equation}
H=\frac{1}{2} \left( p_1^2+p_2^2+p_3^2 \right)-\frac{1}{2} \left(
q_1^2+q_2^2+q_3^2-1\right)^2-\frac{\sigma_2^2}{2}
q_2^2-\frac{\sigma_3^2}{2}q_3^2\label{27}
\end{equation}
in Cartesian coordinates. To write the Hamiltonian in elliptic
coordinates, note that for $N=3$ we have:
\[
\alpha=1+\bar{\sigma}_2^2+\bar{\sigma}_3^2\ ,\qquad
\beta=\bar{\sigma}_2^2\bar{\sigma}_3^2+\bar{\sigma}_2^2+\bar{\sigma}_3^2
\]
\[
A(\lambda_a)=(\lambda_a-1)(\lambda_a-\bar{\sigma}_2^2)(\lambda_a-\bar{\sigma}_3^2)
\]
\[
\Lambda'(\lambda_1)=(\lambda_1-\lambda_2)( \lambda_1-\lambda_3);
\Lambda'(\lambda_2)=(\lambda_2-\lambda_1)( \lambda_2-\lambda_3);
\Lambda'(\lambda_3)=(\lambda_3-\lambda_1)( \lambda_3-\lambda_2)
\]
Hence, $H=\displaystyle \sum_{a=1}^3
\frac{1}{\Lambda'(\lambda_a)}H_a$, where
\begin{equation}
H_a=-2 A(\lambda_a) \pi_a^2-\frac{1}{2} \left[
\lambda_a^2(\lambda_a-\bar{\sigma}_2^2)
(\lambda_a-\bar{\sigma}_3^2)\right]\label{28}
\end{equation}

The separatrix trajectories, those in one-to-one correspondence
with solitary waves of kink type in the encompassing field theory,
are fully determined by the equations (\ref{26}) restricted to the
$N=3$ case: \beq C_2= &&\left|
\frac{\sqrt{1-\lambda_1}+\sigma_2}{\sqrt{1-\lambda_1}-
\sigma_2}\right|^{\sigma_3 {\rm sign}(\pi_1)} \cdot \left|
\frac{\sqrt{1-\lambda_1}-\sigma_3}
{\sqrt{1-\lambda_1}+\sigma_3}\right|^{\sigma_2 {\rm
sign}(\pi_1)}\cdot\nonumber \\ && \left|
\frac{\sqrt{1-\lambda_2}+\sigma_2}{\sqrt{1-\lambda_2}-\sigma_2}
\right|^{\sigma_3 {\rm sign}(\pi_2)}\cdot \left|
\frac{\sqrt{1-\lambda_2}-\sigma_3}{\sqrt{1-\lambda_2}+\sigma_3}
\right|^{\sigma_2 {\rm sign}(\pi_2)}\cdot\label{29}\\ &&\left|
\frac{\sqrt{1-\lambda_3}+\sigma_2}{\sqrt{1-\lambda_3}-
\sigma_2}\right|^{\sigma_3 {\rm sign}(\pi_3)} \cdot \left|
\frac{\sqrt{1-\lambda_3}-
\sigma_3}{\sqrt{1-\lambda_3}+\sigma_3}\right|^{\sigma_2 {\rm
sign}(\pi_3)}\nonumber \eeq
\medskip
\noindent where $C_2={\rm exp} \{ 2\gamma_2 \sigma_2 \sigma_3
(\sigma_2^2-\sigma_3^2) \}$ is constant, and:
\medskip
\beq C_3= &&\left|
\frac{\sqrt{1-\lambda_1}-1}{\sqrt{1-\lambda_1}+1}\right|^{\sigma_2
\sigma_3 (\sigma_2^2-\sigma_3^2){\rm sign}(\pi_1)}\cdot \left|
\frac{\sqrt{1-\lambda_1}+\sigma_2}{\sqrt{1-\lambda_1}-\sigma_2}
\right|^{\sigma_3 \bar{\sigma}_3^2 {\rm
sign}(\pi_1)}\cdot\nonumber\\ && \left|
\frac{\sqrt{1-\lambda_1}-\sigma_3}{\sqrt{1-\lambda_1}+\sigma_3}
\right|^{\sigma_2 \bar{\sigma}_2^2 {\rm sign}(\pi_1)}\cdot \left|
\frac{\sqrt{1-\lambda_2}-1}{\sqrt{1-\lambda_2}+1}\right|^{\sigma_2
\sigma_3 (\sigma_2^2-\sigma_3^2){\rm sign}(\pi_2)}\cdot\nonumber\\
&& \left|
\frac{\sqrt{1-\lambda_2}+\sigma_2}{\sqrt{1-\lambda_2}-\sigma_2}
\right|^{\sigma_3 \bar{\sigma}_3^2 {\rm sign}(\pi_2)}\cdot \left|
\frac{\sqrt{1-\lambda_2}-\sigma_3}{\sqrt{1-\lambda_2}+\sigma_3}
\right|^{\sigma_2 \bar{\sigma}_2^2 {\rm
sign}(\pi_2)}\cdot\label{30}\\ &&\left|
\frac{\sqrt{1-\lambda_3}-1}{\sqrt{1-\lambda_3}+1}\right|^{\sigma_2
\sigma_3 (\sigma_2^2-\sigma_3^2){\rm sign}(\pi_3)}\cdot \left|
\frac{\sqrt{1-\lambda_3}+\sigma_2}{\sqrt{1-\lambda_3}-\sigma_2}
\right|^{\sigma_3 \bar{\sigma}_3^2 {\rm
sign}(\pi_3)}\cdot\nonumber \\ && \left|
\frac{\sqrt{1-\lambda_3}-\sigma_3}{\sqrt{1-\lambda_3}+\sigma_3}
\right|^{\sigma_2 \bar{\sigma}_2^2 {\rm sign}(\pi_3)}\nonumber
\eeq
\medskip
\noindent with $C_3={\rm exp} \{ 2\gamma_3 \sigma_2 \sigma_3
\bar{\sigma}_2^2\bar{\sigma}_3^2 (\sigma_2^2-\sigma_3^2) \}$.

Integration of (\ref{25}) in the $N=3$ case shows the time-table
of the particle in each trajectory, or, the kink form factor:
\medskip
\beq C_1(\tau)\, =&&  \left|
\frac{\sqrt{1-\lambda_1}-\sigma_2}{\sqrt{1-\lambda_1}+\sigma_2}
\right|^{\sigma_3\bar{\sigma}_2^2 {\rm sign}(\pi_1)}\cdot \left|
\frac{\sqrt{1-\lambda_1}+\sigma_3}{\sqrt{1-\lambda_1}-\sigma_3}
\right|^{\sigma_2\bar{\sigma}_3^2 {\rm
sign}(\pi_1)}\cdot\nonumber\\ && \left|
\frac{\sqrt{1-\lambda_2}+\sigma_2}{\sqrt{1-\lambda_2}-\sigma_2}
\right|^{\sigma_3 \bar{\sigma}_2^2 {\rm sign}(\pi_2)}\cdot \left|
\frac{\sqrt{1-\lambda_2}-\sigma_3}{\sqrt{1-\lambda_2}+\sigma_3}
\right|^{\sigma_2\bar{\sigma}_3^2 {\rm sign}(\pi_2)}
\cdot\label{31}\\ && \left|
\frac{\sqrt{1-\lambda_3}-\sigma_2}{\sqrt{1-\lambda_3}+\sigma_2}
\right|^{\sigma_3 \bar{\sigma}_2^2 {\rm sign}(\pi_3)}\cdot \left|
\frac{\sqrt{1-\lambda_3}+\sigma_3}{\sqrt{1-\lambda_3}-\sigma_3}
\right|^{\sigma_2 \bar{\sigma}_3^2 {\rm sign}(\pi_3)}\nonumber
\eeq
\medskip

\noindent if $C_1(\tau)= {\rm exp}\{
2(\gamma_1+\tau)(\sigma_3^2-\sigma_2^2)\sigma_2\sigma_3 \}$.
Therefore, there is a family of kinks parametrized by the
integration constants $\gamma_2$, $\gamma_3$: it corresponds to
the family of curves in $P_3(\infty)$ determined by the
intersection of the surfaces defined by (\ref{29}) and (\ref{30}).
The third constant, $\gamma_1$, fixes the center of the kink, the
point where the energy density reaches its maximum value.

Better intuition of the kink shapes requires  an interpretation of
the solutions described by equations (\ref{29}) and (\ref{30}) in
Cartesian coordinates. We shall describe how is this achieved in
the next sub-sections, but before this it is convenient to note
some details of the change of coordinates from Cartesian to
elliptic in ${\bf R}^3$:
\begin{eqnarray}
q_1^2&=& \frac{1}{\sigma_2^2\sigma_3^2}
(1-\lambda_1)(1-\lambda_2)(1-\lambda_3)\nonumber \\ q_2^2&=&
\frac{-1}{\sigma_2^2 (\sigma_3^2-\sigma_2^2)}
(\bar{\sigma}_2^2-\lambda_1)
(\bar{\sigma}_2^2-\lambda_2)(\bar{\sigma}_2^2-\lambda_3)\label{32}\\
q_3^2 &=& \frac{-1}{\sigma_3^2 (\sigma_2^2-\sigma_3^2)}
(\bar{\sigma}_3^2-\lambda_1)
(\bar{\sigma}_3^2-\lambda_2)(\bar{\sigma}_3^2-\lambda_3)\nonumber
\end{eqnarray}

The change of coordinates is singular at the three ${\bf R}^2$
coordinate planes; $q_1=0$, $q_2=0$ and $q_3=0$. The image of the
$q_1=0$ plane is a unique face, $\lambda_3=1$, of the
$P_3(\infty)$ parallelepiped:
\begin{equation}
-\infty<\lambda_1 \leq \bar{\sigma}_3^2 \leq \lambda_2 \leq
\bar{\sigma}_2^2 \leq\lambda_3\leq1\label{33a}
\end{equation}

The $q_2=0$ plane, however, is mapped into faces
$\lambda_2=\bar{\sigma}_2^2$ and $\lambda_3=\bar{\sigma}_2^2$,
while the $q_3=0$ plane goes to faces $\lambda_2=\bar{\sigma}_3^2$
and $\lambda_1=\bar{\sigma}_3^2$ of $P_3(\infty)$. Observe that
$g_{11}(\bar{\sigma}_3^2,\lambda_2,\lambda_3)=
g_{22}(\lambda_1,\bar{\sigma}_3^2,\lambda_3)=
g_{22}(\lambda_1,\bar{\sigma}_2^2, \lambda_3)=
g_{33}(\lambda_1,\lambda_2,\bar{\sigma}_2^2)=
g_{33}(\lambda_1,\lambda_2,1)=\infty$. The whole ${\bf R}^3$ space
is mapped in $P_3(\infty)$. Due to the symmetry under the group
$G={\bf Z}_2\times {\bf Z}_2\times {\bf Z}_2$ generated by $q_a\to
- q_a$, the mapping (\ref{32}) is eight to one in regular points
of ${\bf R}^3$: to any point in the interior of $P_3(\infty)$
correspond eight points in ${\bf R}^3$ away from the coordinate
planes. These planes are fixed loci of some subgroup of $G$.

The asymptotic conditions (\ref{3}) in $q_a$  restrict
 the motion to the compact
sub-space $D^3$ of ${\bf R}^3$ bounded by the tri-axial ellipsoid:
\begin{equation}
q_1^2+\frac{q_2^2}{\bar{\sigma}_2^2}+\frac{q_3^2}{\bar{\sigma}_3^2}=1\label{33}
\end{equation}
and are satisfied by finite action and zero energy trajectories.
Elliptic coordinates are best suited for demonstrating such a
restriction. In this coordinate system $D^3$ is mapped to the
finite parallelepiped $P_3(0)$:
\begin{equation}
0\leq \lambda_1 \leq \bar{\sigma}_3^2 \leq \lambda_2 \leq
\bar{\sigma}_2^2 \leq \lambda_3\leq 1\label{34}
\end{equation}
The unique non-singular face of $P_3(0)$ with respect to the
change of coordinates is $\lambda_1=0$ and the inverse image of
this face is the ellipsoid (\ref{33}). The asymptotic conditions
(\ref{3}) force $\eta_b=0, \forall b$, and thus, by (\ref{22}),
$H_a=0, \forall a$, for the finite action solutions. $\lambda_2$
and $\lambda_3$ are bounded, see (\ref{34}). Thus, we focus on,
\begin{equation}
H_1=0\Rightarrow \frac{1}{2}\pi_1^2+\frac{1}{8}\frac{\lambda_1^2}{\lambda_1-1}=0\label{100}
\end{equation}
Equation (\ref{100}) describes the motion of a particle with zero
energy moving under the influence of a potential
\begin{eqnarray}
{\cal
V}(\lambda_1)&=&\frac{1}{4}\frac{\lambda_1^2}{\lambda_1-1},\hspace{1cm}
 -\infty<\lambda_1\leq {\bar \sigma}_3^2\nonumber\\&=&
 \infty,\hspace{2,2cm}
  {\bar \sigma}_3^2<\lambda_1<\infty\nonumber
\end{eqnarray}
${\cal V}(\lambda_1)$ has a maximum at $\lambda_1=0$ and goes to
$-\infty$ when $\lambda_1$ tends to $-\infty$; therefore, bounded
motion occurs only in the $\lambda_1\in [0,{\bar \sigma}_3^2]$
interval and the trajectories giving rise to kinks lie in
$P_3(0)$, seen in elliptic coordinates, or $D^3$ in Cartesian
space.

In Figure 1 the whole picture is depicted and we notice the
following important elements of the dynamics:

\begin{figure}[htbp]
\begin{center}
\epsfig{file=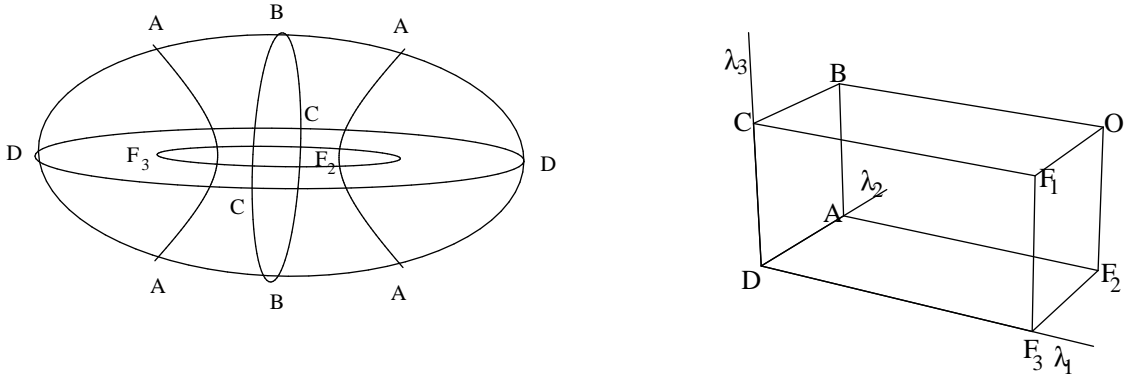,height=5cm}
\end{center}
\caption{\small (a) The domain $D^3$ in $\bf {R^3}$: Cartesian
coordinates (b) The domain $P_3(0)$ in $\bf {R^3}$: Jacobi
elliptic coordinates}
\end{figure}

\noindent - Points: (1) the origin. This is a fixed point of
$G={\bf Z}^{\times 3}$ and thus only one point O in $D^3$ is
mapped to the vertex O in $P_3(\infty)$. (2) Points B, C, D: these
are the intersection points of the three distinguished ellipses ,
$q_1^2+\frac{q_2^2}{\bar{\sigma}_2^2}=1$,
$q_1^2+\frac{q_3^2}{\bar{\sigma}_3^2}=1$ and
$\frac{q_2^2}{\bar{\sigma}_2^2}+
\frac{q_3^2}{\bar{\sigma}_3^2}=1$, in the ellipsoid (\ref{33}).
They are fixed points under the action of a sub-group ${\bf
Z}_2^{\times 2}$ of $G$ and thus, two points in the boundary of
$D^3$ are mapped to a single point in the boundary of $P_3(0)$. D
is the point where the two vacuum points $\vec{v}^{\pm}$ are
mapped and hence it is a very  important point of the dynamics:
every finite action trajectory starts and ends at D. (3) Points
F$_1$, F$_2$, F$_3$, the foci of the above ellipses. Again two
points in $D^3$ are mapped in a unique point in $P_3(0)$. (4) The
umbilicus A of the ellipsoid $\lambda_1=0$ is another
characteristic point; in Cartesian coordinates A corresponds to
four points in the boundary of $D^3$ because they are invariant
only under a ${\bf Z}_2$ sub-group of $G$.
\medskip

\noindent - Curves:
\begin{itemize}
\item The ellipse with foci F$_2$
\begin{equation}
\frac{q_1^2}{\sigma_3^2}+\frac{q_2^2}{\sigma_3^2-\sigma_2^2}=1\label{35}
\end{equation}
in the $q_3=0$ plane passing through F$_3$ and F$_1$. This is the
edge $\lambda_2=\bar{\sigma}_3^2=\lambda_1$ in $P_3(0)$. Observe
that four points on the ellipse (\ref{35}) are mapped to one point
in the edge of $P_3(0)$, because it is invariant under a ${\bf
Z}_2$ sub-group of $G$. The map leading to F$_3$ and F$_1$ is,
however, two to one: the invariance group of these points is
bigger, ${\bf Z}_2^{\times 2}\subset G$.

\item The hyperbola
\begin{equation}
\frac{q_1^2}{\sigma_2^2}-\frac{q_3^2}{\sigma_3^2-\sigma_2^2}=1\label{36}
\end{equation}
in the $q_2=0$ plane passing through F$_2$ and A and having foci
F$_3$ (the edge $\lambda_2=\bar{\sigma}_2^2=\lambda_3$ in
$P_3(0)$).
\end{itemize}

The above points and curves play a special r\^ole in the
definition of the elliptic coordinates and are also \lq\lq
critical loci" of the dynamics.

\subsection{Generic Kinks}

The generic kinks of the $N=3$ system are the trajectories given
by the solutions of (\ref{29})-(\ref{30}) for non-zero finite
values of $C_2$ and $C_3$. The solutions of the implicit equations
(\ref{29})-(\ref{30}) cannot be graphically represented by means
of the built-in functions of Mathematica. We use a numerical
algorithm implemented in Mathematica to obtain the graphic
portrait of the trajectories. The algorithm allows us to calculate
an arbitrary number of points on the orbit. These points joined by
straight segments provide a visualization of the trajectory. There
is a special step and an iteration of routine steps in the
procedure, which is based on the Newton-Raphson method.

\noindent \underline{First step.} Identification of two points on the trajectory.

For given values of $C_2$, $C_3$, $\sigma_2$, $\sigma_3$ and a
choice of signs, we set the first variable to the \lq\lq point"
$\lambda_1=\bar{\lambda}_1$. (\ref{29})-(\ref{30}) becomes a
system of two equations in two unknowns that can be solved by the
Newton-Raphson method with starting values
$(\lambda_2^0,\lambda_3^0)$. The outcome is a point
$P_1\equiv(\bar{\lambda}_1,\bar{\lambda}_2,\bar{\lambda}_3)$ on
the trajectory. We repeat this operation starting from
$\lambda_1=\bar{\lambda_1}+\epsilon=\bar{\lambda}_1'$ to find a
second point
$P_2\equiv(\bar{\lambda}_1',\bar{\lambda}_2',\bar{\lambda}_3')$ on
the orbit.

$\bar{\lambda}_1$, $\bar{\lambda}_1'$, $\lambda_2^0$ and
$\lambda_3^0$ are chosen at random; good convergence is attained
if these points belong to the middle zones of the variation ranges
of $\lambda_1$, $\lambda_2$ and $\lambda_3$ or, at least, they are
far away from the singularities on the faces of $P_3(0)$

\noindent \underline{Successive steps.}

$P_1$ and $P_2$ provide an approximation of the curve by the
secant line joining them. For some small $\delta \in {\bf R}^+$,
we choose $P_3^0=P_1+\delta (P_2-P_1)$ as the starting value of
the Newton-Raphson procedure applied to the solution of equations
(\ref{29}) and (\ref{30}); we thus obtain the point $P_3$ on the
curve. $P_2$ and $P_3$ lead to guesstimate by the same token
another value $P_4^0$ that produce the next point $P_4$ on the
orbit and now the iteration is obvious. Replacing $\delta$ by
$-\delta$ we travel along the opposite sense on the trajectory.
The algorithm stops when one of the three variables $\lambda_1$,
$\lambda_2$, $\lambda_3$ reaches its extreme value; it is applied
independently on each stage, determined by the signs of $\pi_a$
and the global trajectory is obtained by the demand for
continuity.

We now describe the portrait of these orbits. Having fixed
$\gamma_2$ and $\gamma_3$, the corresponding kink trajectory is a
non-plane curve in the interior of $P_3(0)$ that starts from the
vacuum point D, reaches the top face BCF$_1$O and hits the edge
AF$_2$. It then goes to the edge F$_1$F$_3$, back again to the top
face, hits the edge AF$_2$ a second time, the top face a third
time and ends at D: see Figure 2 and Figure 3. Varying $\gamma_2$
and $\gamma_3$ in the range of finite real numbers, other similar
trajectories are obtained that hit the edges AF$_2$ and F$_1$F$_3$
at different points. Given a sense of time there therefore exists
a two-parameter family of kink trajectories in one-to-one
correspondence with the points in the interior of AF$_2$ and
F$_1$F$_3$. It should be mentioned that a whole congruence of
trajectories parametrized by the interior of F$_1$F$_3$ converges
at one single point in the interior of AF$_2$ and viceversa.

\begin{figure}[htbp]
\begin{center}
\epsfig{file=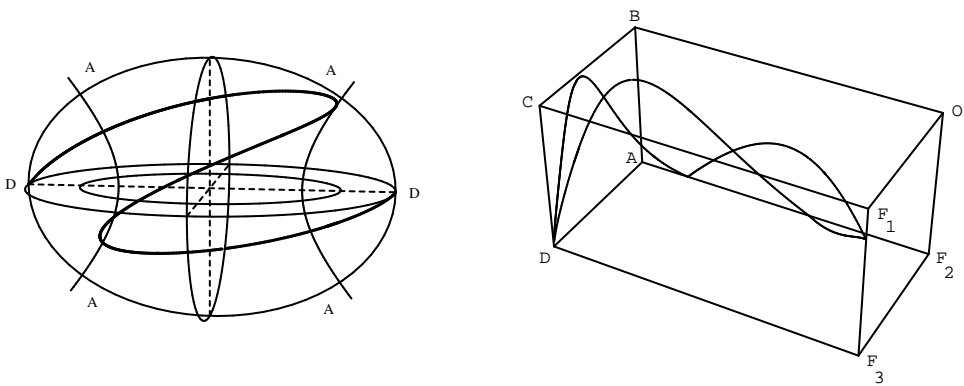,height=5cm}
\end{center}
\caption{\small A generic kink drawn both in $D^3$ and in
$P_3(0)$. Observe in $D^3$ that the generic kink is a heteroclinic
trajectory  }
\end{figure}

\begin{figure}[htbp]
\begin{center}
\epsfig{file=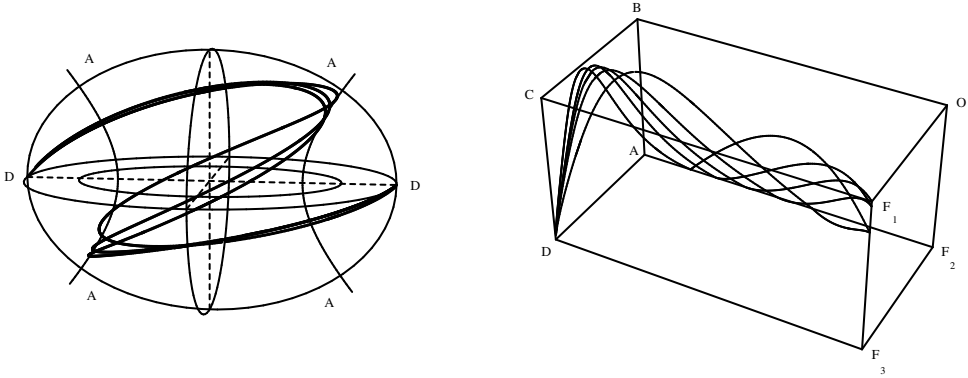,height=5cm}
\end{center}
\caption{\small Several generic kinks, all of them intersecting
once with the edge F$_1$F$_3$ and twice with the edge AF$_2$ of
$P_3(0)$.}
\end{figure}

The translation of a generic kink trajectory to Cartesian
coordinates is a delicate matter; due to the non-uniqueness of the
mapping implied by the change of coordinates special care is
necessary in the analysis of the trajectory near the special
conics (\ref{35})-(\ref{36}) where several options are a priori
possible. Since it requires continuity and derivability to the
trajectories in the interior of the ellipsoid (\ref{33}), the
interior of $D^3$, the behaviour of the curve is mostly fixed. Two
choices, the specification of D$=\vec{v}^-$ as the starting point
and the location of the intersection of the kink trajectory with
the $q_1=0$ plane in the quadrant characterized by $q_2>0$,
$q_3>0$, completely fix the itinerary.

There is a crossroad where the \lq\lq particle" touches the
$q_1>0$ branch of the hyperbola (\ref{36}) and turns back towards
the ellipse (\ref{35}). There, the movement enters the $q_3<0$
half-space and the kink trajectory reaches the other branch,
$q_1<0$, of the critical hyperbola after crossing the $q_2=0$
plane. At this stage, the particle makes its way for a third
crossing of the $q_2=0$ plane and, finally, the journey ends at
D$=\vec{v}^+$. This kind of kink trajectory is therefore
heteroclinic: it starts and ends at different  unstable points, so
that:
\begin{equation}
Q_1^T=\frac{1}{2} \int_{-\infty}^\infty d\tau
\frac{dq_1}{d\tau}=\pm 1,\quad  Q_2^T=Q_3^T=0\label{37}
\end{equation}

We call these topological kinks TK3 because they have three
non-null components:
\[
q_1(\tau)=\phi_1(x)\neq 0,\quad q_2(\tau)=\phi_2(x)\neq 0,\quad
q_3(\tau)=\phi_3(x)\neq 0
\]

It should be noted that a unique, apparently non-derivable, kink
trajectory in elliptic coordinates corresponds to eight derivable
trajectories in Cartesian coordinates: the choices of $\vec{v}^-$
or $\vec{v}^+$ and $q_3<0$ or $q_3>0$, $q_2<0$ or $q_2>0$ as the
starting point and initial quadrant give the eight possibilities.

The energy of a three-component topological kink is the action of
the trajectory times $\frac{m^3}{\lambda^2\sqrt{2}}$ and hence
computable from formula (\ref{23}) for the $N=3$ case:

\begin{eqnarray}
\frac{\lambda^2 \sqrt{2}}{m^3}E_{\rm
TK3}&=&\int_0^{\bar{\sigma}_3^2}
\frac{\lambda_1d\lambda_1}{\sqrt{1-\lambda_1}}
+\int_{\bar{\sigma}_3^2}^{\bar{\sigma}_2^2}
\frac{2\lambda_2d\lambda_2}{\sqrt{1-\lambda_2}} +
\int_{\bar{\sigma}_2^2}^1\frac{3\lambda_3
d\lambda_3}{\sqrt{1-\lambda_3}}\label{38}\\ &=& \frac{4}{3}
+\frac{2}{3} \left[ \sigma_3
(3-\sigma_3^2)+\sigma_2(3-\sigma_2^2)\right]\nonumber
\end{eqnarray}

It is independent of $\gamma_2$ and $\gamma_3$ and hence the same
for every kink in the TK3 family.

\subsection{Enveloping Kinks}

There is another family of $N=3$ kinks living on the surface
$M_3\equiv \left\{ (\lambda_1, \lambda_2, \lambda_3) /\right. $
$\left. \lambda_1=0 \right\}$, the unique face of $P_3(0)$ where
the elliptic coordinates are not singular. In $M_3$, the
Hamiltonian becomes:
\begin{equation}
H=\sum_{a=2}^3 \frac{H_a}{\Lambda'(\lambda_a)}=\sum_{a=2}^3
\left\{ -\frac{2A(\lambda_a)}{\Lambda'(\lambda_a)}\pi_a^2
-\frac{1}{2} \frac{\lambda_a^2(\lambda_a-\bar{\sigma}_2^2)
(\lambda_a-\bar{\sigma}_3^2)}{\Lambda'(\lambda_a)}\right\}
\label{39}
\end{equation}
and therefore there is a two-dimensional system hidden inside the
$N=3$ model which is Hamilton-Jacobi separable. The orbit
equations,
\begin{eqnarray}
C&=& \left| \frac{\sqrt{1-\lambda_2}-
\sigma_2}{\sqrt{1-\lambda_2}+\sigma_2}\right|^{\sigma_3 {\rm
sign}(\pi_2)}\cdot \left|
\frac{\sqrt{1-\lambda_2}+\sigma_3}{\sqrt{1-
\lambda_2}-\sigma_3}\right|^{\sigma_2 {\rm sign}(\pi_2)}\cdot
\nonumber\\ && \left|
\frac{\sqrt{1-\lambda_3}-\sigma_2}{\sqrt{1-\lambda_3}+
\sigma_2}\right|^{\sigma_3 {\rm sign}(\pi_3)}\cdot \left|
\frac{\sqrt{1-\lambda_3}+\sigma_3}{\sqrt{1-\lambda_3}-
\sigma_3}\right|^{\sigma_2 {\rm sign}(\pi_3)} \label{40}
\end{eqnarray}
are parametrized by only one real constant $\gamma_2$ ($C=
e^{\sigma_2\sigma_3 (\sigma_3^2-\sigma_2^2)\gamma_2}$).

\begin{figure}[htbp]
\begin{center}
\epsfig{file=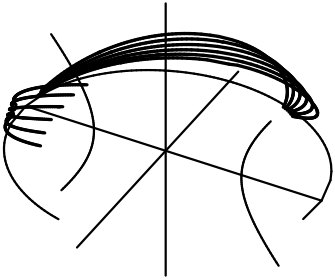,height=5cm}
\end{center}
\caption{\small The NTK3 family}
\end{figure}

The Mathematica plot of these solutions is shown in Figure 4.
Having fixed $\gamma_2$, the corresponding kink trajectory is a
plane curve in $M_3$ that starts from the vacuum point D, reaches
the top edge BC, goes to the umbilicus A and then back to the edge
BC, to end finally in the vacuum point D. The value of $\gamma_2$
determines the points in BC where the trajectory bounces back and
thus the one-parameter family of this kind of kink trajectories is
in one-to-one correspondence with the points in the interior of
BC.

In Cartesian coordinates the enveloping kinks are trajectories
that unfold on the ellipsoid (\ref{33}):
\[
q_1^2+\frac{q_2^2}{\bar{\sigma}_2^2}+\frac{q_3^2}{\bar{\sigma}_3^2}=1
\]

The starting point is either D$=\vec{v}^+$ or D$=\vec{v}^-$ and
the trajectories also end in either D$=\vec{v}^+$ or
D$=\vec{v}^-$. Associated with \lq\lq homoclinic" trajectories,
the corresponding kinks are \lq\lq non-topological":
$Q_1^T=Q_2^T=Q_3^T=0$. The three Cartesian components $q_a$ differ
from zero and the appropriate name for this kind of solitary wave
is a non-topological kink of three components, NTK3 for short.
Every NTK3 trajectory on its way from D$=\vec{v}^{\pm}$ to
D$=\vec{v}^{\pm}$ crosses the umbilicus point of the ellipsoid.
Note that, again, eight trajectories in the Cartesian space ${\bf
R}^3$ correspond to one trajectory in $P_3(0)$: the particle has
the freedom to choose the points $\vec{v}^+$ or $\vec{v}^-$ as
base points of the curve. Having fixed one of them, the trajectory
may develop in the half-ellipsoids determined in (\ref{37}) by
$q_3\leq 0$ or $q_3\geq 0$ and, finally, there are two travelling
senses in each orbit.

Also, the energy of a three-component non-topological kink is
essentially the action of the NTK3 trajectory:
\begin{equation}
\frac{\lambda^2 \sqrt{2}}{m^3} E_{\rm NTK3}=
\int_{\bar{\sigma}_3^2}^{\bar{\sigma}_2^2}\frac{\lambda_2d\lambda_2}{\sqrt{1-
\lambda_2}}+
\int_{\bar{\sigma}_2^2}^1\frac{2\lambda_3d\lambda_3}{\sqrt{1-
\lambda_3}}=2\left( \sigma_2+\sigma_3-
\frac{\sigma_2^2+\sigma_3^2}{3}\right)\label{38b}
\end{equation}
according to the Hamilton-Jacobi theory.

\subsection{Embedded Kinks}

Three-component topological and non-topological kinks arise as
genuine solitary waves in the $N=3$ model. Restriction to the
$q_3=0$ and/or $q_2=0$ planes shows that the $N=2$ system is
included twice, once in each plane, in the $N=3$ model. Therefore,
all the solitary waves of the $N=2$ model are embedded twice as
kinks of the larger $N=3$ system. The embedded kinks live on the
$q_2=0$ and $q_3=0$ planes, i.e. the faces of $P_3(0)$ where the
elliptic coordinate system is singular.

\subsubsection*{I. Embedded kinks in the $q_2=0$ plane}

Both $\lambda_2=\bar{\sigma}_2^2$ and $\lambda_3=\bar{\sigma}_2^2$
give $q_3=0$, see (\ref{32}), and hence this coordinate plane in
${\bf R}^3$ is the union of the two faces,
$\lambda_2=\bar{\sigma}_2^2$ and $\lambda_3=\bar{\sigma}_2^2$, of
$P_3(0)$. Therefore, in
\[
M_{2_{\sigma_3}}=\left\{
(\lambda_1,\lambda_2,\lambda_3)/\lambda_3=\bar{\sigma}_2^2\right\}
\sqcup \left\{
(\lambda_1,\lambda_2,\lambda_3)/\lambda_2=\bar{\sigma}_2^2\right\}
= M_{2_{\sigma_3}}^1 \sqcup M_{2_{\sigma_3}}^2
\]
we expect to find all the kinks of the $N=2$ case.

\begin{itemize}

\item In $M_{2_{\sigma_3}}^1$, $\lambda_3=\bar{\sigma}_2^2$, we are in the face of $P_3(0)$ such that $0<\lambda_1<\bar{\sigma}_3^2<\lambda_2<\bar{\sigma}_2^2$, and
\[
q_1^2=\frac{1}{\sigma_3^2}(1-\lambda_1)(1-\lambda_2)\  ,\qquad
q_3^2=\frac{-1}{\sigma_3^2}(\bar{\sigma}_3^2-\lambda_1)(
\bar{\sigma}_3^2-\lambda_2^2)
\]

The Hamiltonian also reduces to the $N=2$ Hamiltonian
\begin{eqnarray*}
H&=&
-\frac{2A(\lambda_1)}{\Lambda'(\lambda_1)}\pi_1^2-\frac{1}{2\Lambda'(\lambda_1)}
\left( \lambda_1^2(\lambda_1-\bar{\sigma}_2^2)(\lambda_1-
\bar{\sigma}_3^2)\right)\\ &&
-\frac{2A(\lambda_2)}{\Lambda'(\lambda_2)}\pi_2^2-\frac{1}{2\Lambda'(\lambda_2)}
\left( \lambda_2^2(\lambda_2-\bar{\sigma}_2^2)(\lambda_2-
\bar{\sigma}_3^2)\right)
\end{eqnarray*}
and the Hamilton-Jacobi method prescribes the equation
\begin{eqnarray}
e^{2\sigma_3 \bar{\sigma}_3^2 \gamma_2}&=&\left( \left|
\frac{\sqrt{1-\lambda_1}-\sigma_3}{\sqrt{1-\lambda_1}+\sigma_3
}\right|\cdot\left| \frac{\sqrt{1-\lambda_1}+1}{\sqrt{1-\lambda
_1}-1}\right|^{\sigma_3} \right)^{{\rm
sign}(\pi_1)}\cdot\nonumber\\ &&\left( \left|
\frac{\sqrt{1-\lambda_2}-\sigma_3}{\sqrt{1-\lambda_2}+\sigma_3
}\right|\cdot\left| \frac{\sqrt{1-\lambda_2
}+1}{\sqrt{1-\lambda_2}-1}\right|^{\sigma_3} \right)^{{\rm
sign}(\pi_2)}\label{39b}
\end{eqnarray}
as ruling the portion of the trajectories at this face, bounded by
the edges AD, AF$_2$, F$_2$F$_3$ and F$_3$D (see Figure 5).

\begin{figure}[htbp]
\begin{center}
\epsfig{file=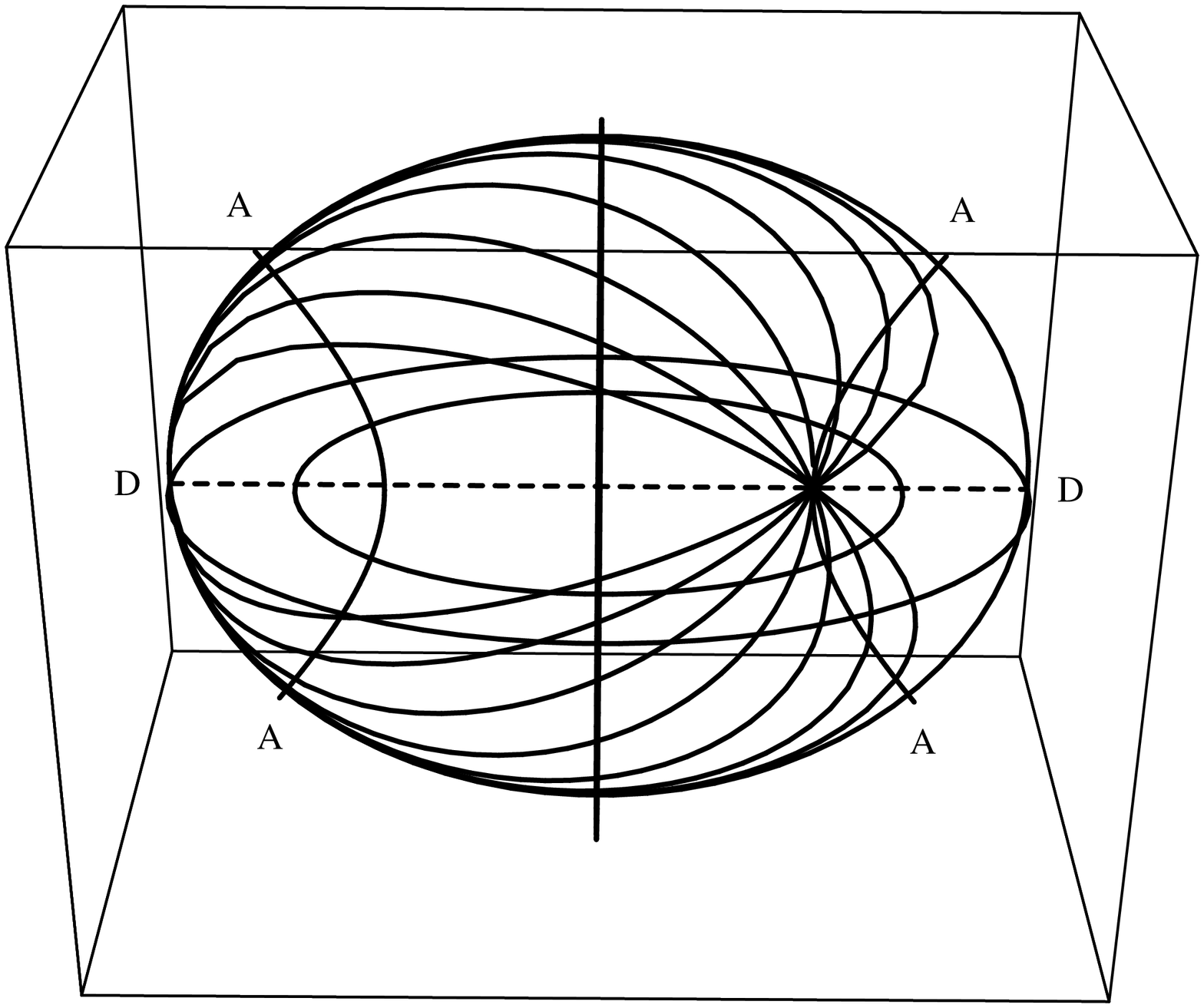,height=5cm}\qquad
\epsfig{file=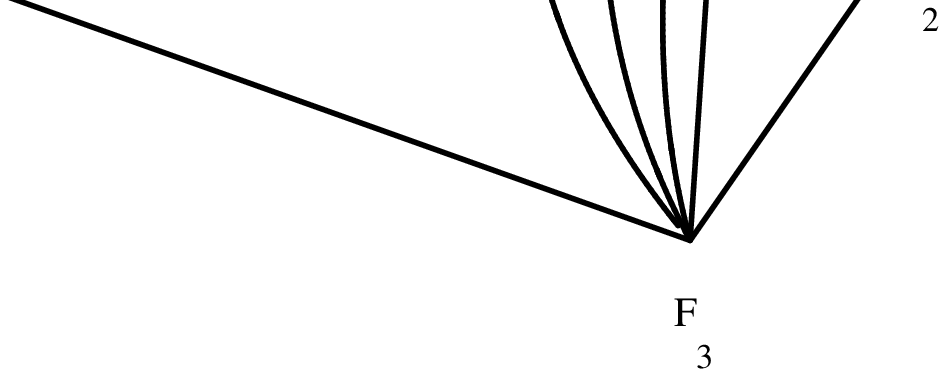,height=5cm}

\end{center}
\caption{\small Several $NTK2_{\sigma_3}$ kink trajectories of the
$N=2$ system embedded in the $q_2=0$ plane }
\end{figure}

\item In $M_{2_{\sigma_3}}^2$, $\lambda_2=\bar{\sigma}_2^2$ and the face in the boundary of $P_3(0)$ is $0<\lambda_1<\bar{\sigma}_3^2<\bar{\sigma}_2^2<\lambda_3<1$. The trajectory equations at this face are:
\begin{eqnarray}
e^{2\sigma_3 \bar{\sigma}_3^2 \gamma_2}&=&\left( \left|
\frac{\sqrt{1-\lambda_1}-\sigma_3}{\sqrt{1-\lambda_1}+\sigma_3
}\right|\cdot\left| \frac{\sqrt{1-\lambda_1}+1}{\sqrt{1-\lambda
_1}-1}\right|^{\sigma_3} \right)^{{\rm sign}(\pi_1)}\cdot\nonumber
\\ && \left( \left|
\frac{\sqrt{1-\lambda_3}-\sigma_3}{\sqrt{1-\lambda_3}+\sigma_3
}\right|\cdot\left| \frac{\sqrt{1-\lambda_3
}+1}{\sqrt{1-\lambda_3}-1}\right|^{\sigma_3} \right)^{{\rm
sign}(\pi_3)}\label{40b}
\end{eqnarray}
and the boundary is formed by the edges AF$_2$, F$_2$O, OB and BA.

\end{itemize}

For finite values of $\gamma_2$, some kink trajectories given by
(\ref{39b})-(\ref{40b}) are depicted in Figure 5. It may be
observed that the trajectory starts at D and then runs through the
face $M_{2_{\sigma_3}}^1$ until the edge AF$_2$. From this point,
the particle enters the $M_{2_{\sigma_3}}^2$ face (here the path
is not derivable), reaches the BO edge and comes back to the
AF$_2$ edge. This is the second point of non-differentiability
re-entering the trajectory the $M_{2_{\sigma_3}}^1$ face. All the
trajectories then meet at the vertex F$_3$ and come back in a
symmetric way to end in the D point. In Cartesian coordinates,
these kink trajectories start and end in either D=$\vec{v}^+$ or
D$=\vec{v}^-$, do not leave the $q_2=0$ plane, and cross either
the focus ($q_1=-\sigma_3$, $q_3=0$) or ($q_1=\sigma_3$, $q_3=0$).
We therefore call them NTK2$_{\sigma_3}$ because they are
two-component non-topological kinks, merely the family of NTK2
kinks of the $N=2$ model, embedded this way within the manifold of
kinks of the $N=3$ system. There are four trajectories of this
kind inside the ellipsoid (\ref{37}) in ${\bf R}^3$ per trajectory
in the boundary of $P_3(0)$: there is freedom to choose
$\vec{v}^+$ or $\vec{v}^-$ and the sense of travel in each orbit.
The NTK2$_{\sigma_3}$ kinks are fixed points of the ${\bf Z}_2$
sub-group of $G={\bf Z}_2^{\times 3}$ generated by $q_2\to -q_2$,
that, however, does not leave invariant the TK3 and the NTK3
trajectories . The energy of these solutions is:
\begin{eqnarray}
\frac{\lambda^2\sqrt{2}}{m^3} E_{{\rm NTK2}_{\sigma_3}}
&=&\int_0^{\bar{\sigma}_3^2}\frac{\lambda_1d\lambda_1}{\sqrt{1-\lambda_1}}
+ \int_{\bar{\sigma}_3^2}^{\bar{\sigma}_2^2}
\frac{2\lambda_2d\lambda_2}{\sqrt{1-\lambda_2}}+
\int_{\bar{\sigma}_2^2}^1
\frac{2\lambda_3d\lambda_3}{\sqrt{1-\lambda_3}}\nonumber\\ &=&
\frac{4}{3} + 2 \sigma_3\left(
1-\frac{\sigma_3^2}{3}\right)\label{41}
\end{eqnarray}

There is a limiting case to this family of kinks: a trajectory
along the DA and AB edges and back to D through the same way.
Elliptic coordinates are even more singular on the edges, but the
dynamical system reduces to a one-dimensional Hamiltonian system
which can be integrated analytically. We have a two-step
trajectory:

At the DA edge, $\lambda_1=0$ and $\lambda_3=\bar{\sigma}_2^2$,
the canonical equations (after use of the first integral) reduce
to:
\begin{equation}
\frac{d\lambda_2}{d\tau} = \pm 2
(\lambda_2-\bar{\sigma}_3^2)\sqrt{1-\lambda_2}\label{42}
\end{equation}
with the solution
\begin{equation}
\lambda_1^{\rm TK2_{\sigma_3}}(\tau)=0,\quad \lambda_2^{\rm
TK2_{\sigma_3}}(\tau)=1-\sigma_3^2 \tanh^2 (\sigma_3 \tau),\quad
\lambda_3^{\rm TK2_{\sigma_3}}(\tau)=\bar{\sigma}_2^2\label{43}
\end{equation}
for $\tau\in (-\infty, \frac{-1}{\sigma_3}
\arctanh\frac{\sigma_2}{\sigma_3} ]\sqcup [\frac{1}{\sigma_3}
\arctanh \frac{\sigma_2}{\sigma_3},\infty)$. The second step
occurs on the AB edge, where, again, the canonical equations
reduce to a single differential equation: if $\lambda_1=0$ and
$\lambda_2=\bar{\sigma}_2^2$,
\begin{equation}
\frac{d\lambda_3}{d\tau}=\pm
2(\lambda_3-\bar{\sigma}_3^2)\sqrt{1-\lambda_3}\label{44}
\end{equation}
has the solution
\begin{equation}
\lambda_1^{\rm TK2_{\sigma_3}}(\tau)=0,\quad \lambda_2^{\rm
TK2_{\sigma_3}}(\tau)= \bar{\sigma}_2^2,\quad \lambda_3^{\rm
TK2_{\sigma_3}}(\tau)=1-\sigma_3^2 \tanh^2 (\sigma_3 \tau)
\label{44b}
\end{equation}
for $\tau\in \left[ \frac{-1}{\sigma_3}
\arctanh\frac{\sigma_2}{\sigma_3} , \frac{1}{\sigma_3} \arctanh
\frac{\sigma_2}{\sigma_3}\right]$. The corresponding kinks in
Cartesian coordinates are TK2$_{\sigma_3}$ and TK2$_{\sigma_3}^*$,
the four two-component topological kinks of the $N=2$ model:
\begin{equation}
\left( \begin{array}{c} q_1^{{\rm TK2}_{\sigma_3}}(\tau) \\
q_2^{{\rm TK2}_{\sigma_3}}(\tau)\\ q_3^{{\rm TK2}_{\sigma_3}}
(\tau)\end{array}\right)= \pm \left(
\begin{array}{c} \tanh (\sigma_3 \tau) \\ 0\\ \pm \bar{\sigma}_3
\sech (\sigma_3 \tau) \end{array}\right)\quad \equiv \left(
\begin{array}{c} \phi_1^{{\rm TK2}_{\sigma_3}}(x,t) \\ \phi_2^{{\rm TK2}_{\sigma_3}}(x,t)\\
\phi_3^{{\rm TK2}_{\sigma_3}}(x,t)\end{array}\right)= \pm \left(
\begin{array}{c} \tanh (\sigma_3 x) \\ 0\\ \pm \bar{\sigma}_3
\sech (\sigma_3 x) \end{array}\right)\label{45}
\end{equation}

Thus, the enveloping kinks of the $N=2$ model are also embedded in
the $N=3$ system. The energy for these solutions and their
anti-kinks is:
\begin{equation}
\frac{\lambda^2 \sqrt{2}}{m^3}E_{{\rm
TK2}_{\sigma_3}}=\int_{\bar{\sigma}_3^2}^{\bar{\sigma}_2^2}
\frac{\lambda_2 d\lambda_2}{\sqrt{1-\lambda_2}}+
\int_{\bar{\sigma}_2^2}^1 \frac{\lambda_3
d\lambda_3}{\sqrt{1-\lambda_3}}=2\sigma_3\left(
1-\frac{\sigma_3^2}{3}\right) \label{46}
\end{equation}

In the $q_2=0$ plane there is still one trajectory that is even
more singular: it is a three step path running on the edges
DF$_3$, F$_3$F$_2$, F$_2$O and back to D through the same way. The
canonical equations and its solutions in the three steps are:

\noindent 1. $\lambda_2=\bar{\sigma}_3^2$ and
$\lambda_3=\bar{\sigma}_2^2$.
\[
\frac{d\lambda_1}{d\tau}=\pm 2\lambda_1 \sqrt{1-\lambda_1},\quad
\tau\in (-\infty,-\arctanh \sigma_3]\sqcup [\arctanh
\sigma_3,\infty)
\]
\[
\lambda_1^{\rm TK1}(\tau)=1-\tanh^2\tau,\quad \lambda_2^{\rm
TK1}(\tau)=\bar{\sigma}_3^2, \quad \lambda_3^{\rm
TK1}(\tau)=\bar{\sigma}_2^2
\]

\noindent 2. $\lambda_1=\bar{\sigma}_3^2$ and
$\lambda_3=\bar{\sigma}_2^2$.
\[
\frac{d\lambda_2}{d\tau}=\pm 2\lambda_2 \sqrt{1-\lambda_2},\quad
\tau\in [-\arctanh \sigma_3,\arctanh \sigma_2]\sqcup [\arctanh
\sigma_2, \arctanh \sigma_3]
\]
\[
\lambda_1^{\rm TK1}(\tau)= \bar{\sigma}_3^2,\quad \lambda_2^{\rm
TK1}(\tau)= 1-\tanh^2\tau, \quad \lambda_3^{\rm
TK1}(\tau)=\bar{\sigma}_2^2
\]

\noindent 3. $\lambda_1=\bar{\sigma}_3^2$ and
$\lambda_2=\bar{\sigma}_2^2$.
\[
\frac{d\lambda_3}{d\tau}=\pm 2\lambda_3 \sqrt{1-\lambda_3},\quad
\tau\in [\arctanh \sigma_2,\arctanh \sigma_2]
\]
\[
\lambda_1^{\rm TK1}(\tau)= \bar{\sigma}_3^2,\quad \lambda_2^{\rm
TK1}(\tau)= \bar{\sigma}_2^2, \quad \lambda_3^{\rm TK1}(\tau)=
1-\tanh^2\tau
\]
\medskip

Only one Cartesian component is different from zero:

\begin{equation}
\left( \begin{array}{c} q_1^{\rm TK1}(\tau) \\ q_2^{\rm
TK1}(\tau)\\ q_3^{\rm TK1}(\tau)\end{array}\right)= \left(
\begin{array}{c} \pm \tanh \tau \\ 0\\ 0 \end{array}\right)\quad
\equiv  \left( \begin{array}{c} \phi_1^{\rm
TK1}(x,t) \\ \phi_2^{\rm TK1}(x,t)\\ \phi_3^{\rm
TK1}(x,t)\end{array}\right)= \left( \begin{array}{c} \pm \tanh x
\\ 0\\ 0 \end{array}\right)\label{47}
\end{equation}
and hence the one-component topological kink of the $N=1$ model is
embedded first in the manifold of kinks of the $N=2$ model, and
then in the $N=3$ system. There are two kinks of this kind in
Cartesian coordinates which are mapped in a unique trajectory in
the boundary of $P_3(0)$. The TK1 trajectories are fixed points of
the ${\bf Z}^{\times 2}$ sub-group of $G$ generated by $q_2\to
-q_2$ and $q_3\to -q_3$. The energy is
\begin{equation}
\frac{\lambda^2 \sqrt{2}}{m^3}E_{\rm
TK1}=\int_0^{\bar{\sigma}_3^2}\frac{\lambda_1d\lambda_1}{\sqrt{1-\lambda_1}}+\int_{\bar{\sigma}_3^2}^{\bar{\sigma}_2^2}
\frac{\lambda_2 d\lambda_2}{\sqrt{1-\lambda_2}}+
\int_{\bar{\sigma}_2^2}^1 \frac{\lambda_3
d\lambda_3}{\sqrt{1-\lambda_3}}=\frac{4}{3} \label{48}
\end{equation}

\subsubsection*{Embedded Kinks in the $q_3=0$ plane}

The DF$_3$, F$_3$F$_2$ and F$_2$O edges form the intersection of
the $q_2=0$ and $q_3=0$ planes. Therefore, the TK1 kinks also live
in the $q_3=0$ plane. There is another maximally singular
trajectory living on the \lq\lq edge" in the $q_3=0$ plane:

At the DC edge, $\lambda_2=\bar{\sigma}_3^2$ and $\lambda_1=0$,
the canonical equations for the finite action trajectories are:
\begin{equation}
\frac{d\lambda_3}{d\tau}=\pm
2(\lambda_3-\bar{\sigma}_2^2)\sqrt{1-\lambda_3}\label{49}
\end{equation}

The path
\begin{equation}
\lambda_1^{{\rm TK2}_{\sigma_2}}(\tau)= 0,\quad \lambda_2^{{\rm
TK2}_{\sigma_2}}(\tau)= \bar{\sigma}_3^2, \quad \lambda_3^{{\rm
TK2}_{\sigma_2}}(\tau)=
1-\sigma_2^2\tanh^2(\sigma_2\tau)\label{50}
\end{equation}
solves (\ref{49}) and runs when $\tau$ goes from $-\infty$ to
$+\infty$ from D to D passing through the vertex C at $\tau=0$. In
Cartesian coordinates we recover the four two-component
topological kinks of the $N=2$ model, now embedded in the $q_3=0$
plane:
\begin{equation}
\left( \begin{array}{c} q_1^{{\rm TK2}_{\sigma_2}} (\tau) \\
q_2^{{\rm TK2}_{\sigma_2}} (\tau)\\ q_3^{{\rm TK2}_{\sigma_2}}
(\tau)\end{array}\right)= \pm \left( \begin{array}{c} \tanh
(\sigma_2\tau) \\ \pm \bar{\sigma}_2^2 \sech (\sigma_2 \tau)\\ 0
\end{array}\right)\quad \equiv \left(
\begin{array}{c} \phi_1^{{\rm TK2}_{\sigma_2}} (x,t) \\
\phi_2^{{\rm TK2}_{\sigma_2}} (x,t)\\ \phi_3^{{\rm
TK2}_{\sigma_2}} (x,t)\end{array}\right)= \pm \left(
\begin{array}{c} \tanh (\sigma_2 x) \\ \pm \bar{\sigma}_2^2\sech
(\sigma_2 x)\\ 0 \end{array}\right)\label{51}
\end{equation}

These are heteroclinic trajectories that produce the
TK2$_{\sigma_2}$ and TK2$_{\sigma_2}^*$ topological kinks. The
energy is:
\begin{equation}
\frac{\lambda^2 \sqrt{2}}{m^3}E_{{\rm
TK2}_{\sigma_2}}=\int_{\bar{\sigma}_2^2}^1\frac{\lambda_3
d\lambda_3}{\sqrt{1-\lambda_3}}=2\sigma_2\left(
1-\frac{\sigma_2^2}{3}\right)\label{52}
\end{equation}

\begin{figure}[htbp]
\begin{center}
\epsfig{file=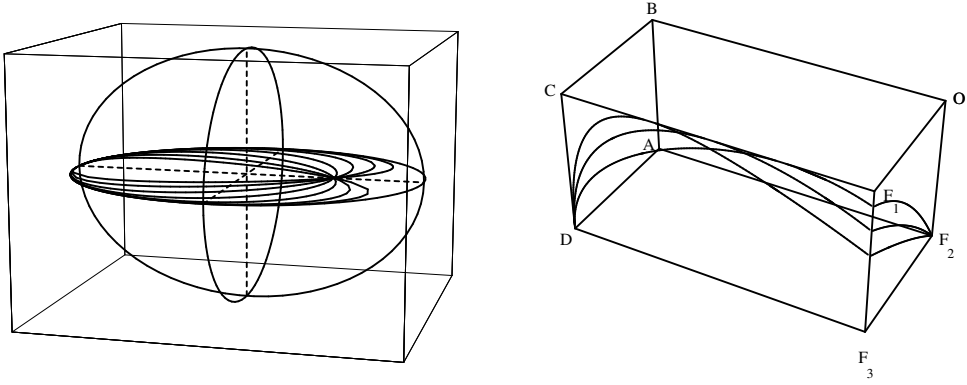,height=5cm}
\end{center}
\caption{\small NTK2$_{\sigma_2}$ kink trajectories of the $N=2$
system embedded in the $q_3=0$ plane}
\end{figure}

Of course,the full manifold of kinks of the $N=2$ model is
embedded in the $q_3=0$ plane  : the set of kinks of the $N=3$
system is completed by the two-component non-topological kinks
living in the $q_3=0$ plane, see Figure 6. The $q_3=0$ plane is
mapped to the union of two faces in the boundary of $P_3(0)$:
\[
M_{2_{\sigma_2}}=\left\{ (\lambda_1, \lambda_2, \lambda_3)/
\lambda_1=\bar{\sigma}_3^2\right\} \sqcup \left\{ (\lambda_1,
\lambda_2, \lambda_3)/ \lambda_2=\bar{\sigma}_3^2\right\} =
M_{2_{\sigma_2}}^1 \sqcup M_{2_{\sigma_2}}^2
\]

\begin{itemize}
\item 1. In $ M_{2_{\sigma_2}}^1$, $\lambda_1=\bar{\sigma}_3^2$ implies $q_3=0$. Therefore:
\begin{eqnarray*}
q_1^2&=& \displaystyle \frac{1}{\sigma_2^2}
(1-\lambda_2)(1-\lambda_3)\\ q_2^2&=& \displaystyle
\frac{-1}{\sigma_2^2} (\bar{\sigma}_2^2-\lambda_2)(
\bar{\sigma}_2^2-\lambda_3)
\end{eqnarray*}
is a well defined change of coordinates in the range
$\bar{\sigma}_3^2<\lambda_2 <\bar{\sigma}_2^2<\lambda_3 < 1$. In
this region, the interior of the ellipse (\ref{35}), the
trajectories providing kinks are given by the equations:
\begin{eqnarray}
e^{2\sigma_2 \bar{\sigma}_2^2 \gamma_2}&=&\left( \left|
\frac{\sqrt{1-\lambda_2}-\sigma_2}{\sqrt{1-\lambda_2}+\sigma_2
}\right|\cdot\left| \frac{\sqrt{1-\lambda_2}+1}{\sqrt{1-\lambda
_2}-1}\right|^{\sigma_2} \right)^{{\rm sign}(\pi_2)}\cdot
\nonumber \\ && \left( \left|
\frac{\sqrt{1-\lambda_3}-\sigma_2}{\sqrt{1-\lambda_3}+\sigma_2
}\right|\cdot\left| \frac{\sqrt{1-\lambda_3
}+1}{\sqrt{1-\lambda_2}-1}\right|^{\sigma_2} \right)^{{\rm
sign}(\pi_3)} \label{I}
\end{eqnarray}

\item 2. In $M_{2_{\sigma_2}}^2$, $\lambda_2=\bar{\sigma}_3^2$ also implies $q_3=0$. In the range $0< \lambda_1<\bar{\sigma}_3^2 < \bar{\sigma}_2^2<\lambda_3<1$ the change of coordinates is defined as
\begin{eqnarray*}
q_1^2&=& \displaystyle \frac{1}{\sigma_2^2}
(1-\lambda_1)(1-\lambda_3) \\ q_1^2&=& \displaystyle
\frac{1}{-\sigma_2^2} (\bar{\sigma}_2^2-\lambda_1)
(\bar{\sigma}_2^2-\lambda_3)
\end{eqnarray*}

The kink trajectories satisfy the equations:
\begin{eqnarray}
e^{2\sigma_2 \bar{\sigma}_2^2 \gamma_2}&=&\left( \left|
\frac{\sqrt{1-\lambda_1}-\sigma_2}{\sqrt{1-\lambda_1}+\sigma_2
}\right|\cdot\left| \frac{\sqrt{1-\lambda_1}+1}{\sqrt{1-\lambda
_1}-1}\right|^{\sigma_2} \right)^{{\rm sign}(\pi_1)}\cdot\nonumber
\\ && \left( \left|
\frac{\sqrt{1-\lambda_3}-\sigma_2}{\sqrt{1-\lambda_3}+\sigma_2
}\right|\cdot\left| \frac{\sqrt{1-\lambda_3
}+1}{\sqrt{1-\lambda_2}-1}\right|^{\sigma_2} \right)^{{\rm
sign}(\pi_3)}\label{II}
\end{eqnarray}

\end{itemize}

The features of this kind of kinks are identical to the
characteristics of the two-component non-topological kinks that
exist in the $q_2=0$ plane. The only difference is that they have
support in the faces $M_{2_{\sigma_2}}^1$ and $M_{2_{\sigma_2}}^2$
instead of $M_{2_{\sigma_3}}^1$ and $M_{2_{\sigma_3}}^2$ and we
therefore call them NTK2$_{\sigma_2}$. They meet at the vertex
F$_2$, and therefore at the foci $(q_1=\pm \sigma_2,0,0)$ in ${\bf
R}^3$; see Figure 6. The energy is:
\begin{eqnarray}
\frac{\lambda^2 \sqrt{2}}{m^3}E_{{\rm NTK2}_{\sigma_2}}&=& 2\left[
\int_0^{\bar{\sigma}_3^2}\frac{\lambda_1d
\lambda_1}{\sqrt{1-\lambda_1}}+
\int_{\bar{\sigma}_3^2}^{\bar{\sigma}_2^2}\frac{\lambda_2d
\lambda_2}{\sqrt{1-\lambda_2}}+
\int_{\bar{\sigma}_2^2}^{1}\frac{\lambda_3d
\lambda_3}{\sqrt{1-\lambda_3}}\right] =\nonumber\\ &=& \frac{4}{3}
+ 2 \sigma_2 \left( 1-\frac{\sigma_2^2}{3}\right)\label{53}
\end{eqnarray}

In sum: the manifold of kinks of the $N=2$ model is embedded twice
in the $N=3$ system, once in the $q_2=0$ plane and other in the
$q_3=0$ plane. They are sewn togehter by the common TK1, embedded
from the $N=1$ model . The embedded kinks fill the gaps left by
the TK3 families of kinks in the interior of the ellipsoid
(\ref{37}) and also develop through the curves left by the NTK3
families on the boundary of $D^3$. $D^3$ is thus a \lq\lq totally"
geodesic manifold with respect to the separatrices between bounded
and unbounded motion in the $N=3$ dynamical system. The NTK3
family form the envelop of the separatrices and the  NTK3 kinks
are themselves separatrices.

\begin{figure}[htbp]
\begin{center}
\epsfig{file=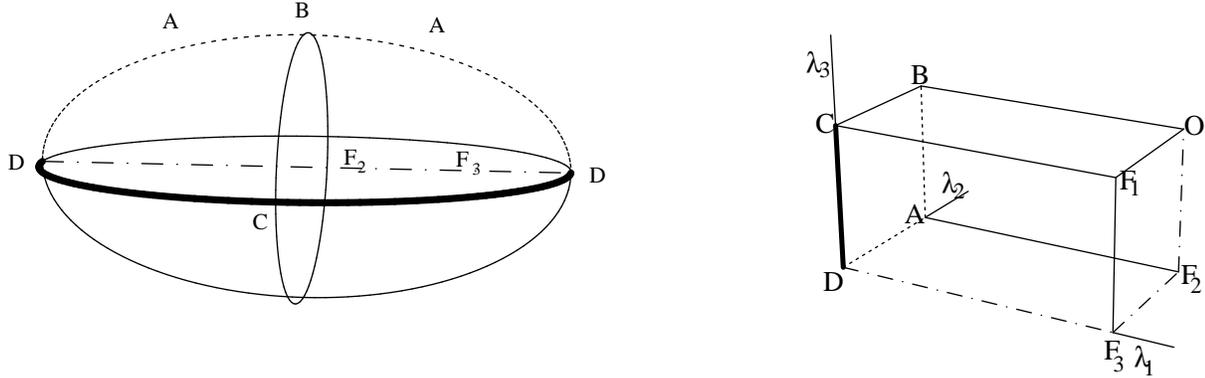,height=5cm}
\end{center}
\caption{\small Plot of the \lq\lq singular" topological kinks TK2$_{\sigma_2}$ (solid line), TK2$_{\sigma_3}$ (broken line) and TK1 (dash-dotted line) in both Cartesian and Elliptic coordinates.}
\end{figure}

\section{Further Comments}

We now infer the general structure of the kink manifold of the
linear $O(N)$-sigma model from the pattern shown by the $O(2)$-
and $O(3)$-sigma models. We can safely state that all the kink
trajectories live in the sub-manifold $D^N \subset {\bf R}^N$
determined by the inequality:
\[
q_1^2+\frac{q_2^2}{\bar\sigma_2^2}+...+
\frac{q_2^2}{\bar\sigma_N^2} \leq 1
\]

There are three categories:
\begin{enumerate}
\item {\bf Generic Kinks}
\begin{itemize}
\item[{\bf A.}] There exists a family of generic kinks parametrized
by $N-1$ real constants that live in the interior of $D^N$. The
intersection loci of the generic kinks are the singular quadrics:
\[
\begin{array}{lcccl}
\frac{q_1^2}{ \bar\sigma_N^2} +
\frac{q_2^2}{\bar\sigma_N^2-\bar\sigma_2^2}+ & \cdots & \cdots \hspace{1cm} +
\frac{q_{N-1}^2}{\bar\sigma_N^2-\bar\sigma_{N-1}^2} & =1, &q_N=0;\  \lambda_1=\lambda_2=\bar{\sigma}_N^2 \\
\frac{q_1^2}{\bar\sigma_{N-1}^2} +
\frac{q_2^2}{\bar\sigma_{N-1}^2-\bar\sigma_2^2}+& \cdots &+
\frac{q_{N-2}^2}{\bar\sigma_{N-1}^2-\bar\sigma_{N-2}^2} -
\frac{q_{N}^2}{\bar\sigma_{N}^2-\bar\sigma_{N-1}^2} & =1, &q_{N-1}=0 ;\  \lambda_2=\lambda_3=\bar{\sigma}_{N-1}^2\\
 &\cdots & \cdots \hspace{1cm} \cdots&  &  \cdots  \\
\frac{q_1^2}{\bar\sigma_{2}^2} -
\frac{q_3^2}{\bar\sigma_{3}^2-\bar\sigma_2^2}-& \cdots & \cdots \hspace{1cm} -
\frac{q_{N}^2}{\bar\sigma_{N}^2-\bar\sigma_{2}^2} & =1, &q_{2}=0 ;\  \lambda_{N-1}=\lambda_N=\bar{\sigma}_2^2
\end{array}
\]
\item[{\bf B.}] Generic kinks are non-topological- hence NTKN- if $N$
even, and topological- hence TKN- if N is odd.
\end{itemize}

\item {\bf Enveloping kinks.}
\begin{itemize}
\item[{\bf A.}] The restriction of the dynamical system to the boundary
$\partial D^N$ of $D^N$, the hyper-ellipsoid:
\[
q_1^2+\frac{q_2^2}{\bar\sigma_2^2}+...+
\frac{q_2^2}{\bar\sigma_N^2} = 1
\]
provides a family of enveloping kinks parametrized by $N-2$ real
constants. Recalling that in elliptic coordinates $\partial D^N$
is characterized by the equation $\lambda_1=0$, the intersection
loci of this congruence are the umbilical sub-manifolds:
\[
\lambda_1=0 \hspace{0.3cm}, \hspace{0.3cm} \lambda_a=\bar\sigma_{N-a+1}^2=
\lambda_{a+1} \hspace{0.2cm}, \hspace{0.2cm} \forall a=2,3,...,N-1
\]
of dimension $N-3$ of the hyper-ellipsoid $\partial D^N$.

\item[{\bf B.}] Enveloping kinks are topological, hence TKN, if $N$ is
even, and non-topological, hence NTKN, if $N$ is odd.
\end{itemize}

\item {\bf Embedded Kinks.}

On the $N-1$ ${\bf R}^{N-1}$ sub-manifolds determined by the
conditions $q_a=0$, if $a=2$ or $3$ or $\dots$ or $N$, the
dynamical system reduces to the mechanical system that arises in
the linear $O(N-1)$-sigma model. Thus, the kink manifold of the
$N-1$ case is included $N-1$ times in the $O(N)$-model, filling
the holes left in the interior of $D^N$ by the generic kinks, and
also covering in $\partial D^N$ the sub-spaces which are not
covereded by the enveloping kinks. Each $N-1$ kink sub-manifold is
not, however, included $(N-1) \times (N-2)$ times in the
$O(N)$-model because the ${\bf R}^{N-2}$ sub-spaces are
intersections of the $N-1$ ${\bf R}^{N-1}$, defined above. The
$N-1$ kink manifolds are not separated but sewn togehter through
the $N-2$ kink sub-manifolds. This is a iterative process in such
a way that the kink manifold of the $O(N-r)$-sigma model is
included $\left(\begin{array}{cc} N-1 \\ r
\end{array}\right)$ times in the kink manifold of the $O(N)$
system.

One can ask what happens if a continuous sub-group $O(r)$ of
$O(N)$ survives as symmetry group of the system. This happens if
the deformation is chosen in such a way that $0<\sigma_2^2=
\sigma_3^2=\cdots=\sigma_r^2<\sigma_{r+1}^2<\dots <\sigma_N^2<1$.
In this case we obtain a sub-manifold of kinks from $O(r)$
rotations around the $q_1$ axis of the kink manifold of the $N=2$
system that lives in the $q_1:q_2$ plane. The remaining kinks
correspond to the solitary waves of the $N=r-1$ system defined in
the orthogonal ${\bf R}^{N-r+1}$ sub-space. Also the deformations
where $1<\sigma_{r+1}^2<\cdots<\sigma_N^2$ are easy to understand.
Finite action trajectories spread out in the domain in ${\bf R}^N$
bounded by the hyper-hyperboloid:
\[
q_1^2+\frac{q_2^2}{\bar\sigma_2^2}+\cdots+\frac{q_r^2}{\bar\sigma_r^2}
-\frac{q_{r+1}^2}{|\bar\sigma_{r+1}^2|}-\cdots
-\frac{q_n^2}{|\bar\sigma_N^2|}=1  .
\]
The kink manifold of this system is the kink manifold of the $N=r$
model defined in the sub-space ${\bf R}^r\subset{\bf R}^N$ such
that $q_{r+1}=\cdots=q_N=0$.

Finally we consider a mild deformation of our model by introducing
asymmetries in the non-harmonic terms of the potential energy and
also adapting the quadratic terms in a suitable manner:
\begin{eqnarray*}
V & = & \frac{1}{2} \left( \phi_1^2 +(1+\varepsilon_2) \phi_2^2+ ...
+(1+\varepsilon_N) \phi_N^2-1 \right)^2+ \\ & & +\frac{\sigma_2^2}{2}
\phi_2^2+... + \frac{\sigma_N^2}{2} \phi_N^2+ \frac{\delta_2}{4}
\phi_2^4+...+ \frac{\delta_N}{4} \phi_N^4
\end{eqnarray*}
The new  non-dimensional constants $\varepsilon_a$ and $\delta_a$
are defined in terms of the old $\sigma_a$'s through:
\[
1+\varepsilon_a=\frac{\sigma_a (\sigma_a+1)}{2}
\hspace{0.1cm},\hspace{0.4cm} \delta_a=\frac{- \sigma_a^3(2+\sigma_a)}{2}
\hspace{0.1cm},\hspace{0.4cm}
a=2,3,...,N
\]

Among the kinks of the deformed linear $O(N)$-sigma model only the
following survive as solitary wave solutions of this perturbed
system:
\begin{enumerate}
\item The TK1.
\[
\phi_1=\tanh x \hspace{0.1cm},\hspace{0.4cm} \phi_2=...=\phi_N=0
\]
\item All the TK2 kinks. On the ellipse,
\[
\phi_1^2+\frac{1+\varepsilon_a}{1-\sigma_a^2} \phi_a^2 =1
\]
the TK2$\sigma_a$ and TK2$^*\sigma_a$ configurations,
\[
\phi_1=\tanh \sigma_a x \hspace{0.1cm},\hspace{0.2cm} \phi_a =
\pm\sqrt{\frac{1-\sigma_a^2}{1+\varepsilon_a}} \, {\rm sech}
\sigma_a x \hspace{0.1cm},\hspace{0.2cm} \phi_b=0
\hspace{0.1cm},\hspace{0.2cm} \forall \, b \neq a, b \neq 1
\]
are solutions of the field equations. The amazing fact is that in
this deformation of the $O(N)$-linear sigma model, discussed by
Bazeia {\it et al.} if $N=2$ \cite{24}, the energy of all these kinks is the
same:
\[
E_{\rm TK1}=E_{{\rm TK2}\sigma_2}=...=E_{{\rm TK2}\sigma_N}=\frac{4}{3 \sqrt{2}}
\frac{m^3}{\lambda^2}
\]
On one hand, we have a deformation of the linear $O(N)$-sigma
model that exhibits a complex variety of kinks; on the other hand,
another deformation of the $O(N)$-model rejects almost every kink
but the simplest ones as solutions, and all of the surviving kinks
are degenerated in energy.
\end{enumerate}
\end{enumerate}
\section{Outlook}

The developments disclosed in this paper suggest a general
strategy in the search for kinks in two space-time dimensional
field theories. When the fields have $N$ components assembled in a
vector representation of the $O(N)$ group, we focus on systems
with symmetry breaking to a discrete sub-group of $O(N)$ which has
more than one element. If the dynamical system that determines the
localized static solutions is completely integrable, all the
solitary waves can be found, at least in principle. Particularly
interesting is the situation where the $N-1$ invariants in
involution with the mechanical energy  act non-trivially on the
manifold of localized solutions and the orbit is a continuous
space. One can then perturb such a system, loosing in the
perturbation many of the solitary wave solutions: only few of the
localized static solutions survive as kinks of the perturbed (more
realistic) model.

We finally list several interesting questions that will be
postponed for future research:

\begin{itemize}

\item study of the structure of the kink manifold of the
deformed linear $O(N)$-sigma model as a moduli space seems to be
worthwhile.

\item A detailed analysis of the sum rules between the energies
of the different kinds of kinks is necessary to fix the structure
mentioned above.

\item A treatment \`a la Bogomolny is also possible. This allows
for a supersymmetric extension of the model in such a way that the
kinks become BPS states.

\item The difficult problem remains of determining the
stability of the different kinds of kinks. Application of the
Morse index theorem helps in finding the stability
properties,which in turn provide information about the
quantization of these topological defects.

\end{itemize}

\section{Acknowledgements}
The authors are grateful to Askold Perelomov for teaching them the
magic of the elliptic Jacobi coordinates and their relationship to
dynamical problems on ellipsoids.

\appendix

\section*{Appendix: Elliptic coordinates}
\addcontentsline{toc}{section}{Appendix: Elliptic Coordinates}

Given any set of $N$ real positive numbers such that
$0<r_1<r_2<\dots <r_N$, let us consider the equation:
\begin{equation}
\sum_{a=1}^N \frac{q_a^2}{r_a-\lambda}=1\label{A1}
\end{equation}

The left-hand member $Q_{\lambda}(\vec{q})=\displaystyle \sum_{a=1}^N
\frac{q_a^2}{r_a-\lambda}$ of this equation can be understood
either as a function of ${\bf R}^N$, for fixed $\lambda\in{\bf
C}$, or as a function of the complex variable $\lambda$, for fixed
$\vec{q}\in {\bf R}^N$. From (\ref{A1}) one immediately
deduces:
\begin{equation}
1-Q_\lambda(\vec{q})=1+\sum_{a=1}^N
\frac{q_a^2}{\lambda-r_a}=\frac{\displaystyle \prod_{a=1}^N
(\lambda-\lambda_a)}{\displaystyle \prod_{a=1}^N
(\lambda-r_a)}=0\label{A2}
\end{equation}

\begin{figure}[htbp]
\begin{center}
\epsfig{file=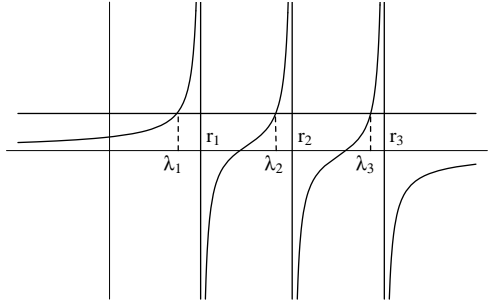,height=4cm}
\end{center}
\caption{\small Plot of the function $y=\displaystyle \sum_{a=1}^3
\frac{q_a^2}{r_a-\lambda}$ and the function $y=1$, see (\ref{A1}),
for fixed values of $q_a$ and $r_a$.}
\end{figure}

Therefore, the $N$ roots $\lambda_a$ of the polynomial in the
numerator of $1-Q_\lambda(\vec{q})$, a rational function of
$\lambda$, are the roots of equation (\ref{A1}). The roots
$\lambda_a$ are also real numbers and $1-Q_\lambda(\vec{q})$ is a
rational function such that $\lambda_1<r_1<\lambda_2<\dots
<r_{N-1}<\lambda_N<r_N$, see Figure 8. To prove this point one
needs to study $1-Q_\lambda(\vec{\rm q})$ along the $\lambda$-real
axis, near the poles $\lambda=r_a$, using Bolzano's theorem.

\noindent {\bf Definition.} The elliptic coordinates of the point
$\vec{q}\equiv (q_1,\dots,q_N)\in {\bf R}^N$ are the roots
$\vec{\lambda}_E\equiv (\lambda_1,\dots,\lambda_N)\in P^N(\infty)$
of $Q_\lambda(\vec{q})=1$.

$P^N(\infty)\subset {\bf R}^N$ is the open sub-space of ${\bf
R}^N$ given by: $-\infty < \lambda_1 < r_1$, $r_1< \lambda_2 <
r_2$, $\dots$, $r_{N-1} < \lambda_N < r_N$. The solution of
(\ref{A1}) for $\lambda=\lambda_1\equiv $constant $\in
(-\infty,r_1)$ is, geometrically, a quadric surface, a
hyperellipsoid, when $\vec{q}$ varies in ${\bf R}^N$. The family
of quadrics obtained by taking $\lambda=\lambda_a\equiv$ constant
$\in (r_{a-1},r_a)$, $a\geq 2$, corresponds to a family of
hyper-hyperboloids of every possible signature in ${\bf R}^N$, if
$Q_{\lambda}(\vec{q})$ is considered as a function of $\vec{q}$.

It is convenient to denote the products in (\ref{A2}) as:
\[
\Lambda(\lambda)=\prod_{a=1}^N (\lambda-\lambda_a)\ ,\qquad
A(\lambda)=\prod_{a=1}^N (\lambda-r_a)
\]
so that
\begin{equation}
1+\sum_{a=1}^N \frac{q_a^2}{\lambda-r_a}=\frac{\displaystyle
\prod_{a=1}^N (\lambda-\lambda_a)}{\displaystyle \prod_{a=1}^N
(\lambda-r_a)}\equiv \frac{\Lambda(\lambda)}{A(\lambda)}\label{A3}
\end{equation}

An explicit formula for defining $q_a^2$ as a function of the
$\lambda$'s , $\forall a$, is obtained by applying the residue
theorem to both members of the equation (\ref{A3}):
\begin{eqnarray}
&& \left. {\rm Res}\left( 1-Q_\lambda(\vec{\rm
q})\right)\right|_{\lambda=r_a}=q_a^2 \nonumber\\ && \label{A4} \\
&& \left. {\rm Res} \left(
\frac{\Lambda(\lambda)}{A(\lambda)}\right)
\right|_{\lambda=r_a}=\frac{\Lambda(r_a)}{A'(r_a)},\quad
A'(r_a)=\left.
\frac{dA(\lambda)}{d\lambda}\right|_{\lambda=r_a}\nonumber
\end{eqnarray}

Therefore:
\begin{equation}
q_a^2=\frac{\Lambda(r_a)}{A'(r_a)}=\frac{\displaystyle
\prod_{b=1}^N (r_a-\lambda_b)}{\displaystyle \prod_{b=1,b\neq a}^N
(r_a-r_b)}\label{A5}
\end{equation}
and we see that the transformation $(q_1,\dots,q_N)\to
(\lambda_1,\dots,\lambda_N)$ is $2^N$ to 1.

Inverting (\ref{A5}) to express $\lambda_a$ as a function of the
$q$'s, $\forall a$, requires one to solve an algebraic equation in
$\lambda_a$ with powers up to $\lambda_a^N$. This is easy for
$N=2$, possible, but very difficult for $N=3,4$ using Cardano's
formulas, and impossible if $N\geq 5$. For this reason, another
derivation of (\ref{A5}) is useful, which in passing allows one to
show identities between Cartesian and elliptic coordinates that
make practical computations possible. To do this, notice that
(\ref{A3}) implies:
\begin{equation}
\Lambda(\lambda)=\prod_{a=1}^N (\lambda-r_a)+\sum_{a=1}^N q_a^2
\prod_{b=1,b\neq a}^N (\lambda-r_b)\label{A6}
\end{equation}

Setting $\lambda=r_c$ in (\ref{A6}) one immediately derives
(\ref{A5}). More important, expanding  the two members of
(\ref{A6}) in a power series in $\lambda$ , we obtain:
\[
\lambda^N+\lambda^{N-1} \left( -\sum_{a=1}^N r_a+\sum_{a=1}^N
q_a^2 \right) +\lambda^{N-2} \left(
\sum\sum_{b<a}^Nr_ar_b-\sum_{a=1}^N q_a^2\sum_{b=1,b\neq
a}^Nr_b\right)+\dots
\]
\[
=\lambda^N-\left( \sum_{a=1}^N \lambda_a\right)
\lambda^{N-1}+\left( \sum\sum_{b<a}^N \lambda_a\lambda_b\right)
\lambda^{N-2}+\dots+(-1)^N \prod_{a=1}^N\lambda_a
\]

Equalizing the coefficients of the terms with the same power of
$\lambda$ in the last equation we have $N$ non trivial identities.
We shall use the equalities between the coefficients of
$\lambda^{N-1}$ and $\lambda^{N-2}$:
\begin{equation}
\sum_{a=1}^N\lambda_a=\sum_{a=1}^N r_a-\sum_{a=1}^N
q_a^2\label{A7}
\end{equation}
\begin{equation}
\sum\sum_{b<a}^N \lambda_a \lambda_b=\sum_{a=1}^N r_a
q_a^2-\sum_{a=1}^N q_a^2 \sum_{b=1}^N r_b+\sum\sum_{b<a}^N r_br_a
\label{A8}
\end{equation}

Another important tool using elliptic coordinates is the Jacobi
lemma:

\noindent {\bf Lemma.}  The expression
\[
\sum_{a=1}^N
\frac{\alpha_a^s}{(\alpha_a-\alpha_1)(\alpha_a-\alpha_2)\dots
\stackrel{(a)}{\dots}\dots (\alpha_a-\alpha_N)}
\]
where $(\alpha_1,\dots,\alpha_N)$ are $N$ real numbers such that
$\alpha_1<\alpha_2<\dots<\alpha_N$ is equal to 0 if $s\leq N-2$
and $1$ for $s=N-1$.

\noindent Proof:

Consider the function
\[
f_s(z)=\frac{z^s}{\displaystyle \prod_{a=1}^N (z-\alpha_a)}
\]
which has $N$ poles in the complex plane at $z=\alpha_a$, $\forall
a$, and another pole at infinity. If $\gamma$ is a closed curve
which is the boundary of a region $D$ of the complex plane
containing all the finite poles of $f_s(z)$, the residue theorem
tells us that:
\[
\frac{1}{2\pi i} \oint_\gamma f_s(z) dz=\sum_{a=1}^N {\rm
Res}(f_s)(\alpha_a)=-{\rm Res}(f_s)(\infty)=-\frac{1}{2\pi
i}\oint_{-\gamma} f_s(z) dz
\]
Also,
\begin{eqnarray*}
&& \sum_{a=1}^N {\rm Res}(f_s)(\alpha_a)=\sum_{a=1}^N
\frac{\alpha_a^s}{(\alpha_a-\alpha_1)(\alpha_a-\alpha_2)\dots
\stackrel{(a)}{\dots}\dots (\alpha_a-\alpha_N)}\\ && {\rm
Res}(f_s)(\infty) = \left\{ \begin{array}{l} 0,\quad \forall
s<N-1\\ -1,\quad s=N-1\end{array} \right.
\end{eqnarray*}
and the lemma is proved.

\medskip

Applying this result to the choice $\alpha_a=\lambda_a$, $\forall
a=1,\dots,N$, we obtain new identities
\begin{eqnarray}
&&\sum_{a=1}^N \frac{\lambda_a^s}{\Lambda'(\lambda_a)}=0,\forall
s<N-1\label{A8b}\\ &&\sum_{a=1}^N
\frac{\lambda_a^{N-1}}{\Lambda'(\lambda_a)}=1\label{A9}
\end{eqnarray}
because
$\Lambda'(\lambda_a)=(\lambda_a-\lambda_1)(\lambda_a-\lambda_2)\dots
\stackrel{(a)}{\dots}\dots (\lambda_a-\lambda_N)$. Alternatively
if we take $\alpha_a=r_a$, the lemma implies that:
\begin{equation}
\sum_{a=1}^N \frac{r_a^s}{A'(r_a)}=0,\forall s<N-1;\qquad
\sum_{a=1}^N \frac{r_a^{N-1}}{A'(r_a)}=1\label{A10}
\end{equation}

An important identity obtained from the lemma is:
\begin{equation}
\sum_{a=1}^N \frac{q_a^2}{(r_a-\lambda_b)(r_a-\lambda_c)}=0,\
\forall b,c\label{A11}
\end{equation}

The normal vectors to the family of quadrics (\ref{A1})
\[
Q_\lambda(\vec{q})=\sum_{a=1}^N \frac{q_a^2}{r_a-\lambda}=1
\]
at the point $\vec{q}\equiv(q_1,\dots,q_N)$, are
\[
\vec{n}(\lambda)\equiv
(n_1(\lambda),\dots,n_N(\lambda))=\left(\frac{q_1}{r_1-\lambda},\dots,\frac{q_N}{r_N-\lambda}\right)
\]

Observe that (\ref{A11}) implies
\[
\sum_{a=1}^N n_a(\lambda_b)n_a(\lambda_c)=0,\  \forall a,b
\]
Therefore, all the quadrics are orthogonal with each other and the
elliptic coordinates form an orthogonal system. The standard
Euclidean metric in Cartesian coordinates can be expressed in
elliptic coordinates in the form:
\[
ds^2=\sum_{a=1}^N dq_a^2=\sum_{a=1}^N\sum_{b=1}^N
g_{ab}(\vec{\lambda}_E) d\lambda_a d\lambda_b
\]
Derivation of the two members of equation (\ref{A5}) leads to:
\[
\frac{2dq_a}{q_a}=(-1)^N\sum_{b=1}^N
\frac{d\lambda_b}{r_a-\lambda_b}
\]
and, using the Jacobi Lemma,
\[
4dq_a^2=q_a^2 \sum_{b=1}^N \frac{d\lambda_b^2}{r_a-\lambda_b}
\]

Finally, we have:
\begin{equation}
g_{aa}=-\frac{1}{4}
\frac{\Lambda'(\lambda_a)}{A(\lambda_a)}=-\frac{1}{4}
\frac{\displaystyle \prod_{b\neq a,b=1}^N
(\lambda_a-\lambda_b)}{\displaystyle \prod_{b=1}^N
(\lambda_a-r_b)},\ g_{ab}=0, \forall a\neq b\label{A12}
\end{equation}

The kinetic energy of a natural dynamical system in elliptic coordinates
is;
\begin{equation}
T=\frac{1}{2} \sum_{a=1}^N\dot{q}_a^2=\frac{1}{2} \sum_{a=1}^N
g_{aa} \dot{\lambda}_a^2\label{A13}
\end{equation}

In terms of the canonical momentum $\pi_a=\frac{\partial
T}{\partial \dot{\lambda}_a}$, $T$ reads:
\begin{equation}
T=\sum_{a=1}^N \pi_a \dot{\lambda}_a-T=\frac{1}{2} \sum_{a=1}^N
\frac{1}{g_{aa}} \pi_a^2=-2 \sum_{a=1}^N
\frac{A(\lambda_a)}{\Lambda'(\lambda_a)}\pi_a^2\label{A14}
\end{equation}

\end{document}